\documentclass[aps,prd,showpacs,singlecolumn,nofootinbib,superscriptaddress,10.5pt]{revtex4-1}
\pdfoutput=1
\usepackage{graphicx,caption,amsmath, amssymb}  
\usepackage{cancel}
\usepackage{color,latexsym,array,multirow,hyperref,verbatim,enumerate}
\usepackage{multirow}
\usepackage{threeparttable}
\usepackage{slashed}
\usepackage{footnote}
\newcommand*{\cmssfont}{\fontfamily{cmss}\selectfont}
\usepackage{soul}

\usepackage{lineno}

\def\PET{\rm p{\!\!\!/}_T}
\def\PEx{\rm p{\!\!\!/}_x}
\def\PEy{\rm p{\!\!\!/}_y}
\def\PEz{\rm p{\!\!\!/}_z}

\def\MET{\rm E{\!\!\!/}_T}

\begin{document}

\title{Search for bottom squarks in the 
baryon-number violating MSSM}

\author{Debjyoti Bardhan}
\email{debjyoti@theory.tifr.res.in}
\affiliation{Department of Theoretical Physics,  
Tata Institute of Fundamental Research,\\
1, Homi Bhabha Road, Mumbai 400005, India}

\author{Amit Chakraborty}
\email{amit@theory.tifr.res.in}
\affiliation{Department of Theoretical Physics,  
Tata Institute of Fundamental Research,\\
1, Homi Bhabha Road, Mumbai 400005, India}

\author{Debajyoti Choudhury}
\email{debajyoti.choudhury@gmail.com}
\affiliation{Department of Physics and Astrophysics,University of Delhi, Delhi 110007, India}

\author{Dilip Kumar Ghosh}
\email{tpdkg@iacs.res.in}
\affiliation{Department of Theoretical Physics, Indian Association for the Cultivation of Science, \\
2A \& 2B, Raja S.C. Mullick Road, Kolkata 700032, India}

\author{Manas Maity}
\email{manas.maity@cern.ch}
\affiliation{Department of Physics, Visva-Bharati, Santiniketan 731235, India}

%
%
\begin{flushright}
{TIFR/TH/16-42}
\end{flushright}
\begin{abstract}
We consider a scenario of Minimal Supersymmetric Standard Model (MSSM)
with $R$-parity violation, where the lightest supersymmetric particle
(LSP) is the lighter sbottom $(\tilde b_1)$. We study the production
of a sbottom pair at the LHC and their subsequent decays through the
baryon number violating (UDD) operators leading to a top pair with
two light quarks.  Looking for both semi-leptonic and fully hadronic
(no leptons) final states, we perform cut-based as well as multivariate
analyses (MVA) to estimate the signal significance at the 13 TeV run
of the LHC. We find that a cut-based analysis can probe sbottom mass upto 
$\sim 750 $ GeV which may be extended upto $\sim 850 $ GeV using MVA with 
$300~{\rm fb}^{-1}$ integrated luminosity.
The fully hadronic final state, however, is not as promising.
\end{abstract}

\pacs{12.60.Jv, 14.65.Ha, 14.65.Bt, 14.80.Ly}

\maketitle

\section{Introduction}
\label{intro}
The discovery of the most elusive particle of the Standard Model (SM),
the Higgs boson, by the LHC during its first run completed the hunt
for the SM particles \cite{Aad:2012tfa,Chatrchyan:2012xdj}.  The SM
has been extremely successful in explaining diverse phenomena 
involving elementary particles. The SM, however,
fails to address many issues, two of the most prominent being the
hierarchy problem and the baryon asymmetry in the universe.
Supersymmetry (SUSY), one of the most popular extensions
of the SM, can not only provide plausible resolutions for these
problems, but can also facilitate answers to other questions
such as the unification of forces, or the presence of dark
matter (DM). Consequently, the ATLAS and CMS collaborations at the LHC have
searched for signatures of the supersymmetric particles. So far,
no signal of any SUSY particles has been seen and this has been
construed to impose severe constraints on the parameter space of the
theory\cite{susy_update}.

One should, however, note that the majority of the SUSY search
strategies at the LHC assume that `R-parity', 
a multiplicative quantum number defined as $R = (-1)^{3B +
  L +2s}$ with $B$, $L$ and $s$ in terms of the baryon number (B),  the 
  lepton number (L) and the spin (S) of the particle, is conserved. Conservation
of R-parity implies that SUSY particles will always be pair produced
and that a heavy SUSY particle will decay into an odd number of
lighter SUSY particles, with or without other SM particles\footnote{In
  most popular models, the decays are into a single lighter SUSY
  particle and one or two (and, only rarely three) SM
  particles.}. This ensures that the lightest SUSY particle (LSP) is
stable. A characteristic signature of an R-parity conserving SUSY
scenario is a final state with large missing transverse energy
($\MET$) due to the presence of the LSP.  Since, in the SM, neutrinos
are the only real sources of missing energy apart from detector acceptance
and resolution effects,
 $\MET$ can be used as a
standard candle to search for these SUSY particles. Besides, in
supersymmetric theories with conserved R-parity, the lightest SUSY
particle, if colourless and electrically neutral, can always act as a
good dark matter candidate.

However, the conservation of R-parity is not guaranteed, and,
if one allows for its violation, an sfermion
can decay to a pair of SM fermions\footnote{Similarly, the gauginos
  and higgsinos would decay into three SM fermions.}, giving rise to
signatures with, at best, only a small missing transverse
energy \cite{Barbier:2004ez}. This negation of one of the standard
features of SUSY searches would, immediately, negate much of the
collider constraints on the SUSY spectrum. Welcome consequences of
this are the easing of fine-tuning on the one
hand~\cite{Carpenter:2006hs}, and, on the other, the possibility of
facilitating electroweak
baryogenesis\cite{Balazs:2004bu,Carena:2002ss,Carena:2008vj} through
the accommodation of a top-squark lighter than what the R-conserving 
scenarios can allow.  From its very definition, one can see that the
violation of
R-parity can be achieved in three ways: violation of either L or B, or 
both. However, if we allow both L and
B to be violated, nothing prevents the proton from
decaying into a meson and a lepton, and, thus, the lower limit on the
proton decay lifetime \cite{proton_decay} places severe constraints on
their products. It is, of course, more natural to ensure proton stability by
insisting on one of the symmetries ($B$ or $L$) being unbroken, and
this is the route that we take. Interestingly, such $R$-violating
scenarios can be easily motivated from supergravity
models~\cite{Allanach:2003eb}. And while violating R-parity implies
that we lose the DM candidate, the dark matter content of the universe
can appear from other sources~\cite{Hui:2016ltb}.

In this paper, we study a R-violating (RPV)
 SUSY scenario in the presence
of baryon number (UDD-type) violating operators alone.  Contrary to
naive expectations, such a scenario can be well-accommodated within a
GUT-framework~\cite{Tamvakis:1996np}, thereby preserving one of the
successes of SUSY. A further ramification is that, unlike in the case
of the $L$-violating couplings, the lack of any excess in the
multilepton channel at the LHC does not impose any worthwhile
constraint on the squark/gluino masses
\cite{ATLAS:2016kjm,ATLAS:2016soo,
  Aaboud:2016hmk,CMS:2016dtq,CMS:2013qda,CMS:swa}. We are faced,
instead, with a multijet
signal\cite{CMS:2015wva,ATLAS:2016yhq,ATLAS:2016sfd,ATLAS:2016nij} and
it has been argued that the large irreducible QCD background would
result in much weaker sensitivity. Performing a collider analysis of
the lightest scalar superpartner of the bottom quark, namely the
sbottom ($\tilde{b}$), subsequently decaying to a top quark and a light down-type quark through
non-zero $\lambda^{\prime\prime}$ couplings, we show that it is not
necessarily so. Depending on the decay of the top quark, the final
state can consist of only hadronic elements (jets), or may contain at
least one lepton. The latter semi-leptonic case is easier to study at
a hadronic collider environment like that of the LHC, since we can tag
on the lepton. Our analysis will take into account the very different
nature of these two possible final states and is thus done in two
parts: first, for a final state with at least one lepton, and second,
for a fully hadronic final state. To study the semi-leptonic final
state, we shall use both the traditional cut-based analysis and
multivariate analysis, while in the hadronic final state, we shall
rely solely on the multivariate analysis.

The rest of the
paper is arranged as follows: in Section II, we briefly introduce
R-parity violating SUSY, noting down the couplings relevant for our
analysis. In Section III, we introduce our simplified model detailing
all the parameters used.  The analysis of a final state with a lepton
is presented in Section IV and in Section V, we perform the analysis
for a completely hadronic final state.  Finally, we conclude in
Section VI.

\section{The R-parity violating MSSM }
In terms of lepton, quark and Higgs superfields one can write down the 
R-parity violating superpotential in the following form \cite{Barbier:2004ez}:  
either bilinear terms or by Yukawa-like trilinear terms. The most generic RPV 
superpotential is given by:
\begin{equation}
W_{\not{R}_p} = \mu_i \hat{H}_u \hat{L}_i 
           + \frac{1}{2} \lambda_{kij} \hat{L}_i \hat{L}_j \hat{E}_k^c 
           + \lambda'_{ijk} \hat{L}_i \hat{Q}_j \hat{D}_k^c 
           + \frac{1}{2} \lambda''_{ijk} \hat{U}_i^c \hat{D}_j^c \hat{D}_k^c 
\label{superpotential}
\end{equation}
where the first three sets of operators violate $L$ while the last set
violates $B$. Here, $i,j $ and $k$ are generation indices, whereas
both $SU(2)$ and $SU(3)$ indices have been suppressed.  Clearly, the
couplings $\lambda_{kij}$ and $\lambda''_{ijk}$ are antisymmetric in
the last two indices and, thus, there are a total of (3 + 9 + 27 =) 39
$L$-- and 9 $B$--violating interactions.  Switching off the first
three sets, and concentrating only on the last, we have, in terms
of the quark and squark fields:
\begin{equation}
\mathcal{L}_{UDD} = -\frac{1}{2}\lambda''_{ijk} \left(\tilde{u}^\star_{iR} \overline{d}_{jR} d^c_{kL} + \tilde{d}^\star_{kR} \overline{u}_{iR} d^c_{jL} + \tilde{d}^\star_{jR} \overline{u}_{iR} d^c_{kL} \right) \ .
\end{equation}
The bounds on the couplings $\lambda^{''}_{ijk}$ are varied. Some of
them are strongly constrained from $n-\bar{n}$
oscillations~\cite{Goity:1994dq} or the LEP data on
$Z$-decays~\cite{Bhattacharyya:1995bw}. The others are only weakly
restricted, for example, through the requirement of their perturbative
under renormalization group flows \cite{Brahmachari:1994wd}.
Compendia of such constraints can be found in
Refs.\cite{Barbier:2004ez,Saha:2002kt,Bhattacharyya:1995pr,deGouvea:2000cf,bhatta_pal,Agashe:1995qm,
  Ghosh:1996bm,Allanach:2003eb,Agashe:2014kda}. It should be noted
that many of the low-energy constraints emanate from effective
four-fermi interactions, and in quoting them a reference squark mass
is used; these bounds need to be scaled appropriately when the
squark mass differs.

As we are interested in the $\tilde b$, one of $j,k$ in
$\lambda''_{ijk}$ must be 3. Similarly, if we demand that the sbottom
should decay into a top, we must have $i = 3$. 
In other words, we are
left with just two choices, namely $\lambda''_{313}$ and
$\lambda''_{323}$, leading to $\tilde b^* \to t + d$ and $\tilde b^*
\to t + s$ respectively.  Since the simultaneous presence of two such
couplings lead to too large a size for flavour changing neutral
currents (FCNC)~\cite{Chakraverty:2000rm, Chakraverty:2000df}, we
assume that only one of the two is non-zero and real.  For the mass
range (of the squarks) that we are interested in, the strongest
constraints are $\lambda''_{313} < 0.1$~\cite{Agashe:2014kda} and
$\lambda''_{323} < 1.89$~\cite{Bhattacharyya:1995bw} respectively.
Even without saturating these bounds, it is obvious that, once
produced, the sbottom may decay promptly, thereby eliminating the
possibility of recognizably displaced vertices 
\cite{Zwane:2015bra,Monteux:2016gag}. We shall assume that while the R-violating couplings are small 
enough to be both consistent with low energy phenomenology as well as 
having at best marginal effect on squark-production, they are large enough to 
prevent displaced vertices, thereby removing tell-tale signatures.

In the presence of `UDD'--type couplings, the decays (direct or
cascades) of squarks and gluinos (the dominantly produced SUSY
particles at the LHC) would, typically, result in multi-jet
configurations with very little missing momentum.  As these are very
difficult to detect (especially in the absence of hard leptons) in the
messy hadronic environment of the LHC, the strong limits on
squark/gluino masses, derived in the context of R-conserving models
(or, even for R-violating, but $B$-conserving ones) do not hold. In
particular, if a pair-produced squark decays directly into a pair of
quarks, the resultant four-jet sample is likely to be overwhelmed by
the QCD background. The situation is ameliorated somewhat if some of
the quarks (rather, the corresponding jets) can be tagged as
this would allow us not only to eliminate much of the background, but
also to use invariant mass combinations to increase the
signal-to-noise ratio. This was used in
Refs.\cite{Choudhury:2005dg,Choudhury:2011ve} and, subsequently, by
the ATLAS collaboration\cite{ATLAS:2016yhq}, to investigate the
pair-production of stops and their decays, through the very same
couplings that we are considering here, to a $b$--quark and a light
quark each. Here, we investigate the complementary
scenario, namely where the sbottom (rather than the stop) is the LSP.

\section{The Simplified Model Spectrum and Simulation} 
 
As we are primarily interested in the lighter bottom squark, we
simplify the spectrum by considering it to be the LSP with the other
SUSY partners being much heavier. In particular, we do not include
gluino production despite the fact that, for similar masses,
$\sigma(\tilde g \tilde g)
\gg \sigma(\tilde b \tilde b^*)$ and that the gluino could easily
decay into a $\bar b \tilde b$ pair, thereby adding to the signal
strength. Indeed, gluino pair-production, with each decaying
into three quarks has been used~\cite{ATLAS:2016nij} to set a limit of
$m_{\tilde g} > 1.08$ TeV and, hence, by making the $\tilde g$ heavy,
we deliberately preclude this contribution altogether.  
Our assumption about the spectrum obviously means that decays through
R-conserving channels are no longer possible and that the sbottom is
forced to decay to two SM quarks with 100\% branching ratio.  
For the choice of the RPV coupling $\lambda''_{313}$ ($\lambda''_{323}$), 
the daughters are the top and a light-quark ($d$ or $s$, as the case may be).
 The top quark can decay either
leptonically or hadronically; thus giving rise to the following final
states :
\begin{itemize}
\item $2 \ell \ell^\prime + b {\bar b} + {\rm jets} + \MET $; 
$\ell, \ell^\prime  = e, \mu $,
\item $1 \ell + b {\bar b} + {\rm jets} + \MET $,
\item $0 \ell + b {\bar b} + {\rm jets} + \MET $.
\end{itemize}
It should be noted that all these channels
will be associated with only a small missing
transverse energy, if any. Final states with multiple jets are
very challenging in the LHC environment and thus require dedicated
studies. Several SM processes which provide similar final state signatures
have been treated as background, particularly, $t \bar{t} +{\rm
jets}$ (upto 2), $t\bar {t}b\bar{b}$, $t\bar {t}Z$, $t\bar {t}W$ and
$t\bar{t}H$, constitute the dominant SM background; QCD multijet events
constitute huge background for the purely hadronic case. 

%
Before we delve into the discussion of signals and backgrounds, let us
examine the parameter space that leads to the spectrum that we
consider. The gaugino mass parameters $M_1$ and $M_2$, as well as the
higgsino mass parameter, $\mu$ are set to 1 TeV, while the value of
$\tan\beta$, the ratio of the vacuum expectation values of the two
Higgs doublets $H^0_u$ and $H^0_d$, is fixed at 10. The masses of the
first two generations of squarks and all the three generations of
sleptons lie around 3 TeV and the mass of the right-handed stop is set
to $\sim$ 1 TeV. The left-handed third generation squark mass is set
to about 1.5 TeV. While the tri-linear couplings $A_t$ is set to $-2$
TeV, the other tri-linear couplings $A_b$ and $A_\tau$ are set to
zero. We also fix the gluino mass parameter ($M_3$) at 2 TeV, while
varying only the right-handed sbottom mass parameter ($m_{b_R}$) to
obtain different sbottom masses. In our analysis we consider six
representative benchmark points with sbottom masses $500$~GeV (BP-1),
$600$~GeV (BP-2), $700$~GeV (BP-3), $800$~GeV (BP-4), $900$~GeV (BP-5)
and $1000$~GeV (BP-6).
\begin{table} [htb]
\begin{tabular}{|l|c|c|c|c|c|c|}
\hline
& BP - 1 & BP - 2 & BP - 3 & BP - 4 & BP - 5 & BP - 6 \\
\hline 
\hline
$m_{\tilde{b}_1}$ (GeV) & 500 & 600 & 700 & 800 & 900 & 1000 \\
\hline
\hline
$\mathcal{B}(B_s \to \mu^+\mu^-)$ & \multicolumn{6}{c|}{$4.27 \times 10^{-9}$}  \\
$\mathcal{B}(b \to s \gamma) $ & \multicolumn{6}{c|}{$3.19 \times 10^{-4}$}  \\
$\Delta m_{B_s} ({\rm ps}^{-1})$ &  \multicolumn{6}{c|}{$18.01$}  \\
$\Delta m_{B_d} ({\rm ps}^{-1})$ & \multicolumn{6}{c|}{$0.403$}  \\
\hline
\end{tabular}
\caption{{\small{\it The first row presents the masses of the bottom squark for the different benchmark points. In the bottom part of the table, the values of the low-energy flavour observables are presented. These remain identical for the different benchmark points.}}}
\label{tab:mass_flav}
\end{table}

The particle spectrum has been generated using SPheno v-3.3.8 \cite{
  Porod:2003um,Porod:2011nf} with the trilinear R-parity violating
model as implemented in SARAH v-4.4.6 \cite{Staub:2008uz,
  Staub:2015kfa}.  FlavorKit \cite{Porod:2014xia} is used to calculate
the low energy flavour observables $b \to s \gamma$ and $B_s \to \mu^+
\mu^-$ and care has been taken to ensure that the benchmark points are
consistent with the flavour physics data \cite{Amhis:2014hma} at
better than 95\% C.L. In particular, the mass differences
$\Delta m_{B_d}$ $\left(\Delta m_{B_s}\right)$ associated with $B_0$--
and $B_s$--mixing (see Table \ref{tab:mass_flav}) are very close to
the experimental measurements \cite{Agashe:2014kda}.  We further
ensure that the 
 spectrum we use at each benchmark point is consistent with
the latest measurements of Higgs mass, Higgs couplings and Higgs
signal strength at the LHC.

It is worth noting that for our analysis, lighter squarks, consistent
with the present bounds, would not
be a problem.  Since the contribution to various flavour processes
from RPV, typically, are
  proportional to $(\lambda''^2/m_{\tilde{q}}^2)$, it is possible to
accommodate a smaller squark mass, provided the couplings are 
reduced accordingly.  In this
scenario, we would, for example, receive 
additional contribution to our signal
events from, say the sstrange. If the sstrange were only slightly
heavier than the sbottom, it would decay to the sbottom along with a
bottom and a strange quark, via an off-shell gluino\footnote{The only 
other channel available to it would be the RPV channel to the top and 
the bottom, which would, again, be largely indistinguishable from
that we consider here.}. Owing to only a
small difference in the masses, the sbottom would be produced almost
at rest with the two other jets being very soft; this would be
indistinguishable from the sbottom pair production scenario and would
thus add to our signal events. We do not consider this, and, thus, the
analysis in this paper is quite conservative.

The signal and background events are generated using {\cmssfont
  MADGRAPH (version 2.2.2)} \cite{Alwall:2014hca}, properly interfaced
with {\cmssfont PYTHIA8 (version 8.210)}
\cite{Sjostrand:2014zea,Sjostrand:2006za} for parton
showering and hadronization. Event sets are then passed through
{\cmssfont DELPHES (version 3.2.0)} \cite{deFavereau:2013fsa} in order
to simulate the detector response. Jets are reconstructed using
{\cmssfont FASTJET (version 3.1.3)} \cite{Cacciari:2011ma}, with $R=
0.4$ using the anti-$k_t$ algorithm \cite{antikt} in the leptonic
case. For the hadronic case, we intend to tag the boosted top quarks
in the final state, which will necessarily be a fat jet; thus, we use
$R = 1.8$ using the C/A algorithm \cite{cambridge_aachen} which is
optimized for tagging moderately boosted tops \cite{Anders:2013oga}.

Jets are selected with $p_{T} >$ 30 GeV and $|\eta| < 2.5$.  Leptons
(electron and muon) are selected with $p_{T} >$ 20 GeV and $|\eta| <
2.4$. To reduce the background contribution of electrons or muons from
semileptonic decays of heavy flavours, a relative isolation criteria is
imposed. The relative isolation parameter, $I_{\rm rel}$, defined
as, 
\begin{eqnarray}
I_{\rm rel} = \frac{\sum_{i\ne P} {p}_T(i)}{p_T (P)}
\end{eqnarray}
with $P$ being the particle of interest (here electron or muon), 
is calculated as the sum of transverse energy of all the 
charged and neutral particles measured in the tracker and calorimeters 
in an isolation cone\footnote{Here, and 
henceforth, $\Delta R\equiv \sqrt{(\Delta\eta)^2 + (\Delta\phi)^2}$ 
is the usual distance measure in the rapidity($\eta$)--azimuthal 
angle ($\phi$) plane.} $\Delta R < 0.3$ around the lepton direction 
divided by the lepton transverse momentum. In our analysis, we 
demand $I_{\rm rel} <$ 0.15.

In the semi-leptonic decays of the top, the final state contains 
multiple leptons and a significant amount of missing transverse energy, 
calculated using 
the $p_T$ of all the visible particles. Our signal topology 
also includes multiple b-jet candidates and in order to tag them as `b-jets', 
we require the angular distance $\Delta R$ between the parton level
b-quark and the jet to be less that 0.4, as implemented in {\cmssfont DELPHES}. 
A $p_T$ dependent b-tagging 
efficiency ($\epsilon_b$) for $|\eta|<2.5$, following the CMS collaboration 
\cite{CMS_btag}, is used to make our analysis more robust: 
\begin{equation}
\epsilon_b = 
\left\{ \begin{array}{ll} 0.75 & {\rm for} \ \ p_T^b \leq 30~{\rm GeV} \\[0.5ex]
                          0.85 & {\rm for} \ \ 30~{\rm GeV} < p_T^b \leq 400~{\rm GeV} \\[0.5ex]
                          0.95 - 0.00025 \, p_T & {\rm for} \ \ 400~{\rm GeV} < p_T^b \leq 800~{\rm GeV} \\[0.5ex]
                          0.65 & {\rm for} \ \ p_T > 800~{\rm GeV}.
               \end{array} \right.   
\end{equation}
Throughout the entire $p_T$ range, following the CMS card,
a mistagging rate of 1\% is assumed for the
 non b-jets.  Note that, the b-tagging
efficiency obtained by the ATLAS collaboration \cite{ATLAS_btag} is
comparable with that of the CMS collaboration.

The cross-section of the $t \bar{t}$ + jets (upto 2)
process is taken from the LHC Top Quark Working Group
\cite{top_working_group}, while that of the $t \bar{t}H$ is taken from
the LHC Higgs Cross-Section Working Group report
\cite{higgs_working_group}.  The NLO cross-section for $t\bar{t}W$ and
$t\bar{t}Z$ are taken from \cite{Maltoni:2015ena}, where the results
have been computed using MSTW2008 parton distribution functions
(PDFs).  We use {\cmssfont PYTHIA} to calculate the cross-section for
the $t\bar{t}b\bar{b}$ process, where the PDF used in the calculation
is CTEQ6L \cite{Pumplin:2002vw} and the factorization scale has been
chosen to be $M_Z$, the mass of the Z-boson. For the signal 
processes, we use the sbottom pair production cross-sections at 
the 13 TeV LHC calculated including the resummation of soft-gluon 
emission at next-to-leading logarithmic accuracy matched to 
next-to-leading order supersymmetric QCD 
corrections \cite{Borschensky:2014cia}.


\section{Leptonic Final State}
In this section we consider the final state in which there is at least
one lepton; thus this analysis includes both leptonic and
semi-leptonic decays of the top quarks.  We first perform a cut-based
analysis on the data sets and then supplement it with a multivariate
analysis.
 
\begin{figure}[!htb]
\includegraphics[scale=0.28]{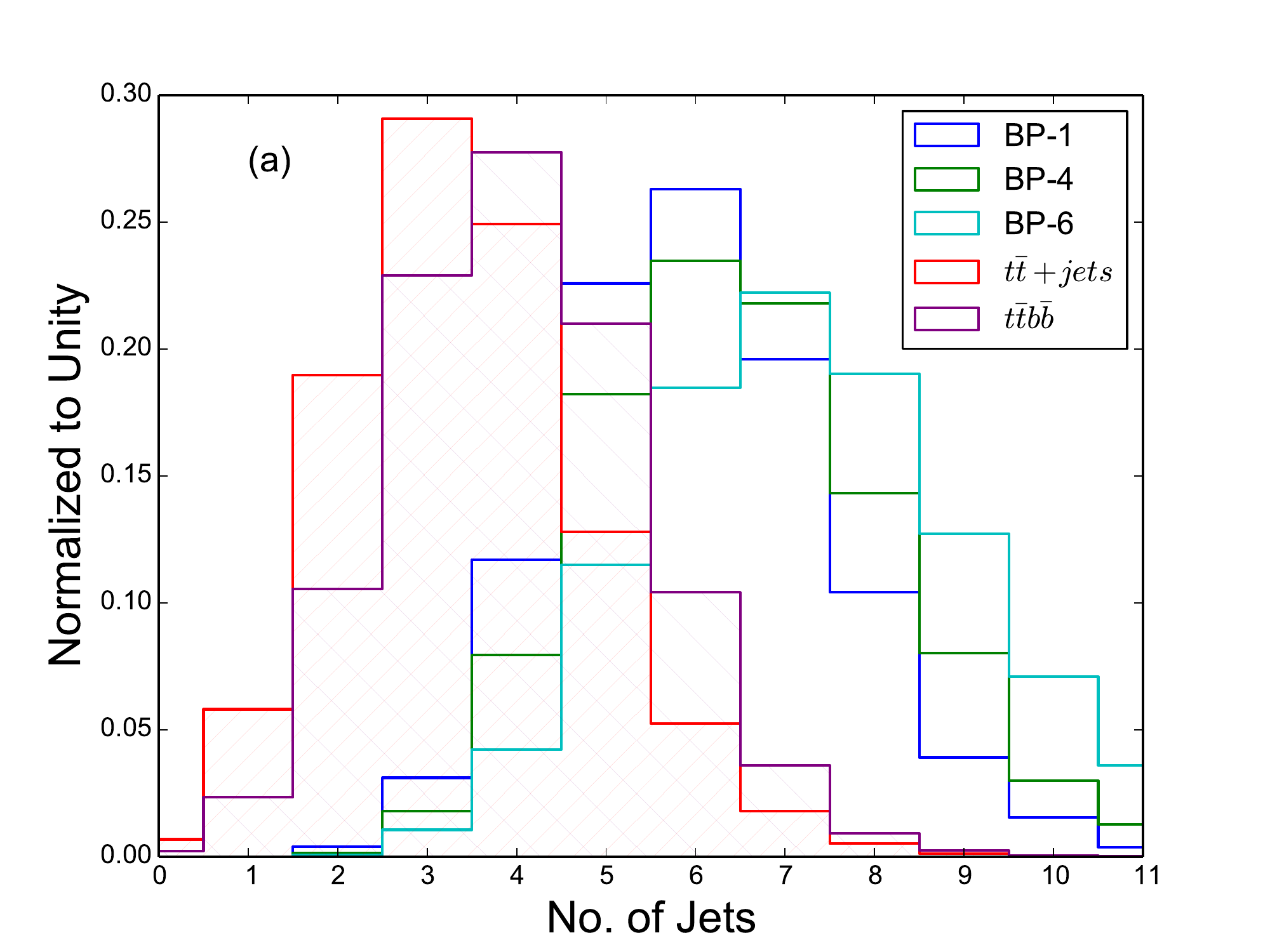}
\includegraphics[scale=0.28]{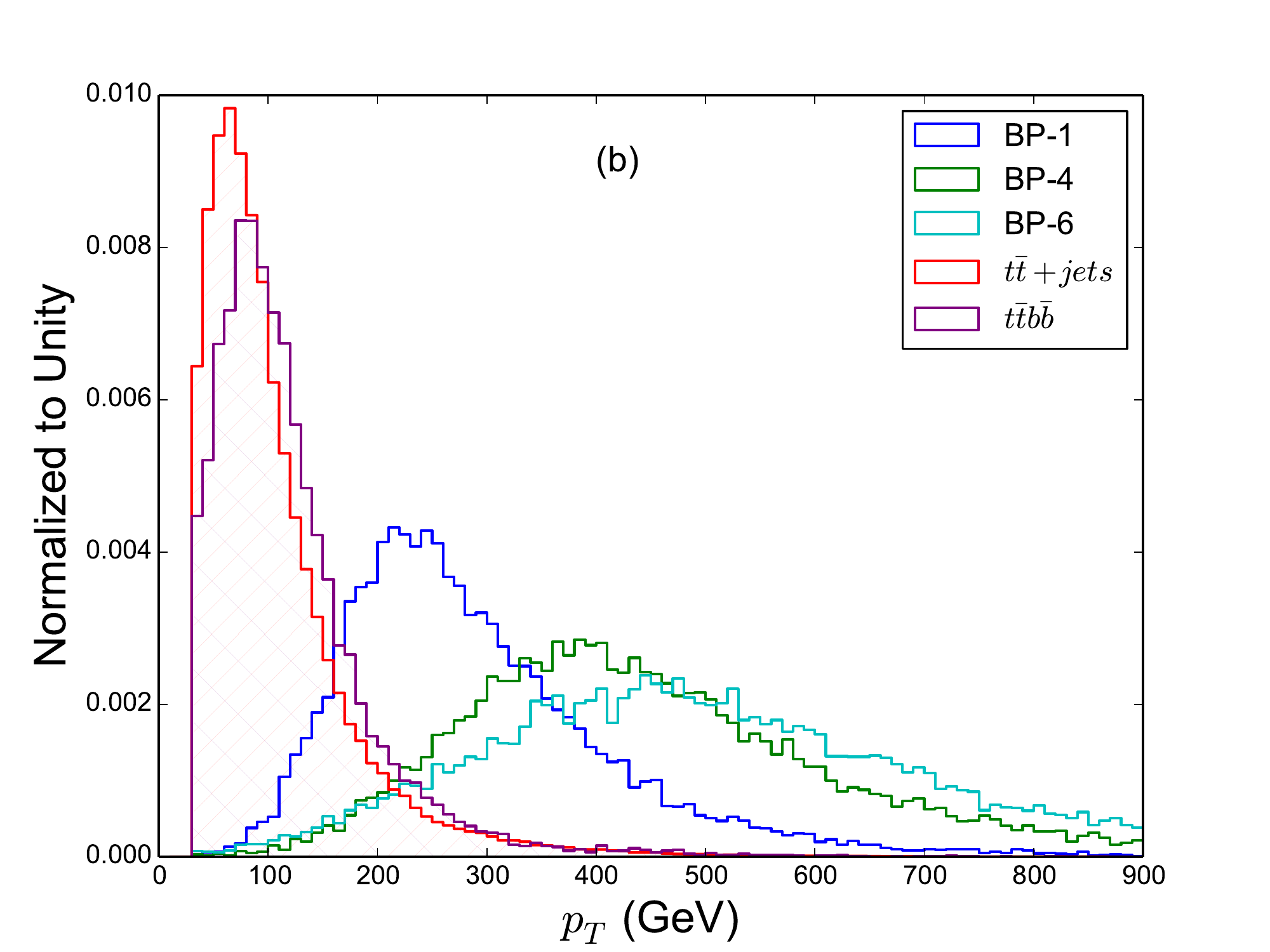}
\includegraphics[scale=0.28]{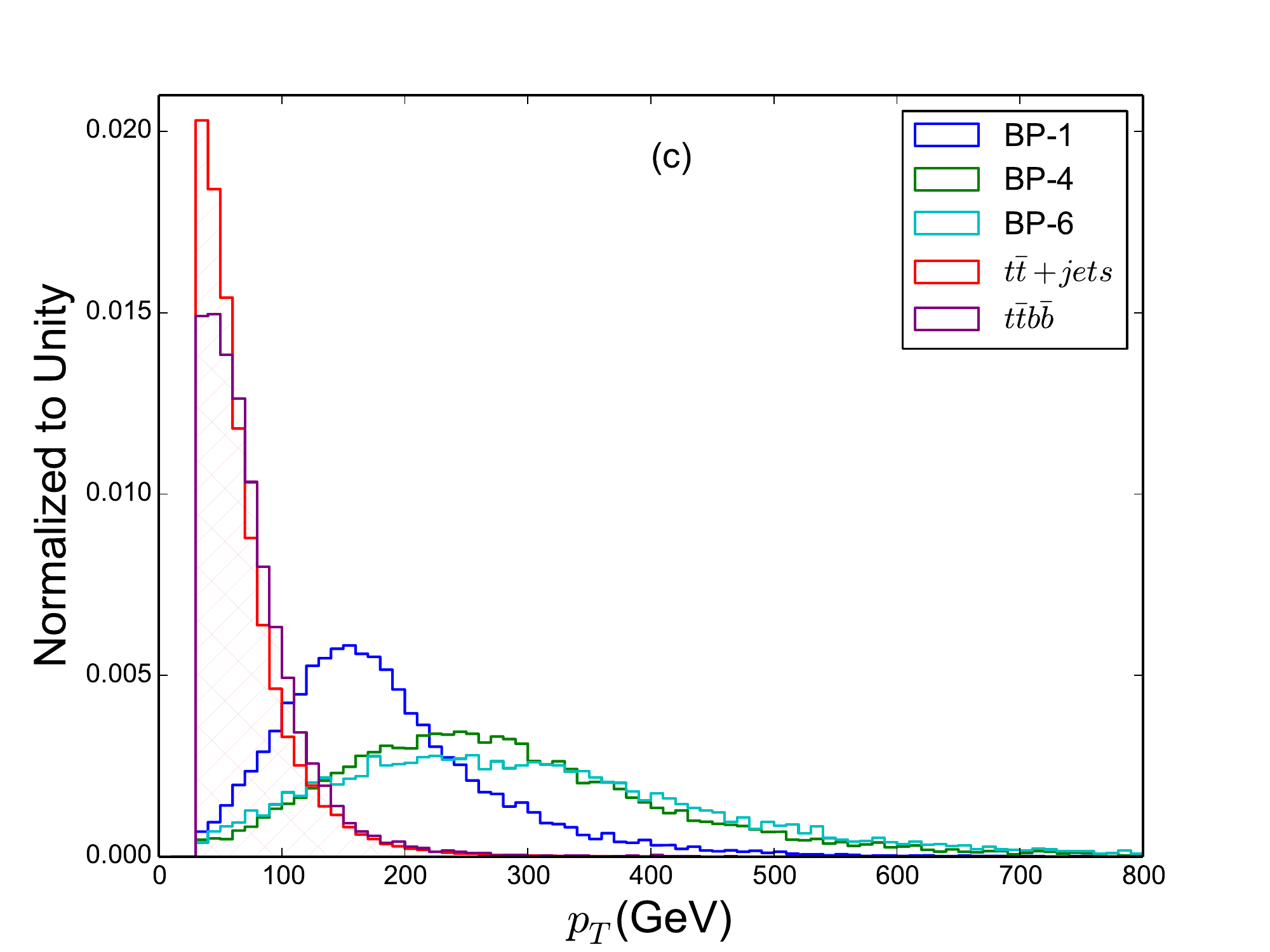}
\caption{\small \it In $(a)$ we show the distribution of the number of
  jets, $(b)$ the $P_T$ distribution of the hardest non b-tagged jet,
  while in $(c)$ the same for the second hardest non b-tagged jet. For
  the sake of clarity, just the two dominant SM background processes
  are shown, viz. $t\bar{t}+{\rm jets}$ and $t\bar{t}b\bar{b}$.}
\label{fig:ptj}
\end{figure}

\begin{figure}[!htb]
\includegraphics[scale=0.4]{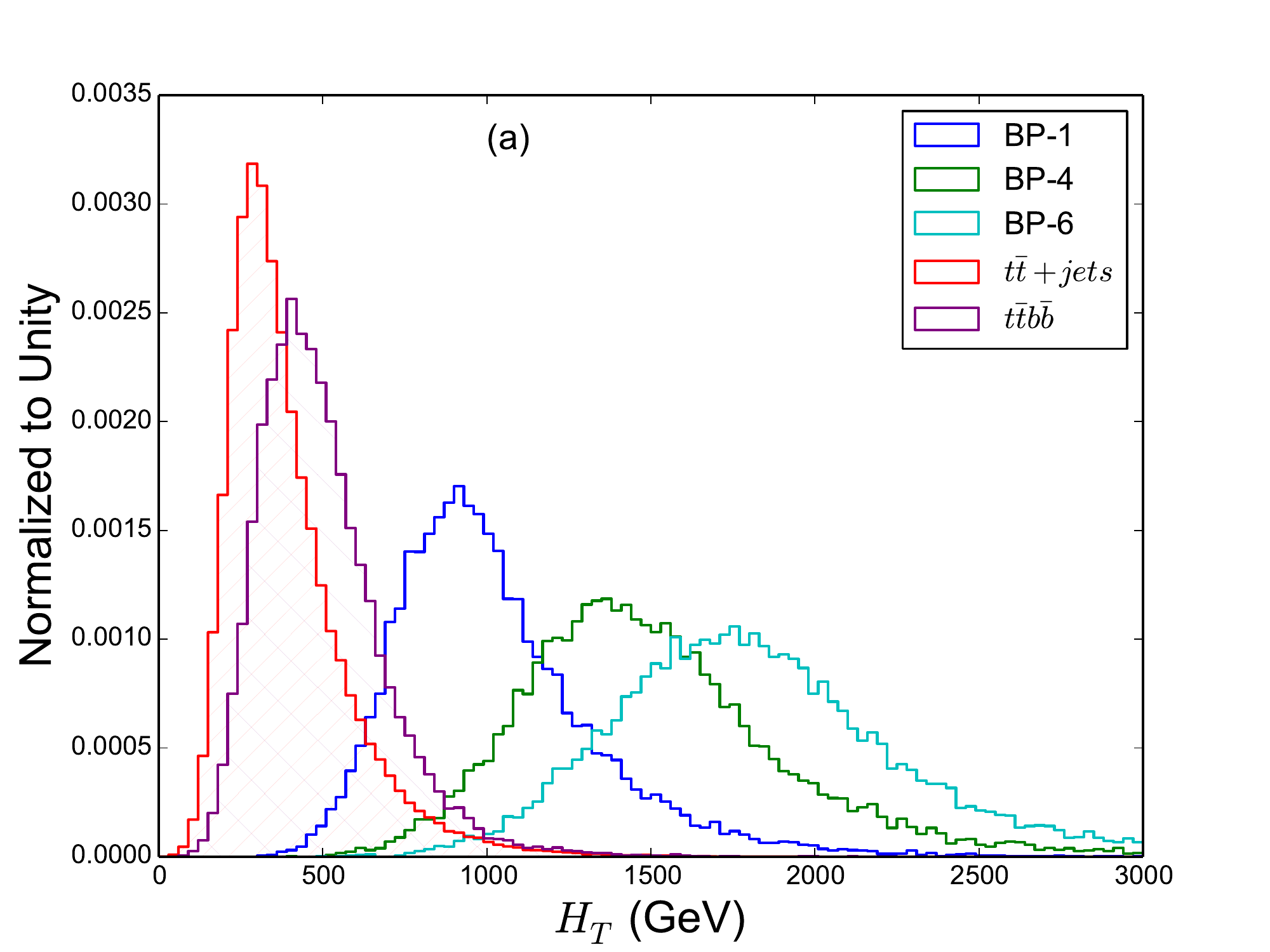}
\includegraphics[scale=0.4]{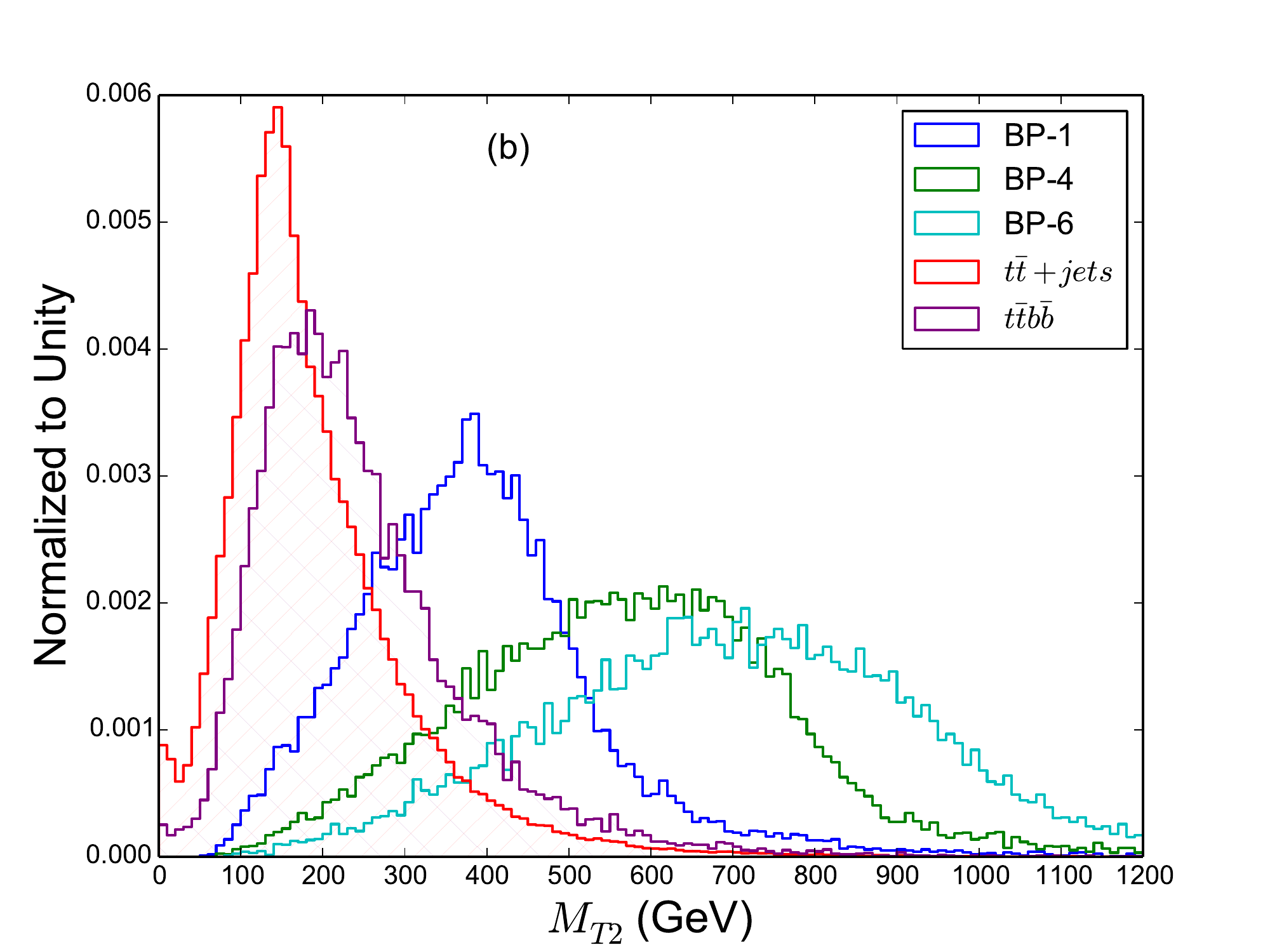}
\caption{\small \it Distribution of the $(a)$ scalar $H_T$ and $(b)$ Stransverse mass variable, $M_{T2}$, 
are displayed.}
\label{fig:HTMt2}
\end{figure}

In Fig.~\ref{fig:ptj}, we present the jet multiplicity and the $p_T$
distribution of the two leading non b-tagged jets. Additionally, in
Fig.~\ref{fig:HTMt2}, the distributions for $H_T$ and $M_{T2}$, both
defined shortly, are also shown. All
the distributions include three representative benchmark points -
BP-1, corresponding to $m_{\tilde{b}_1} = 500$~GeV, BP-4,
corresponding to $m_{\tilde{b}_1} = 800$~GeV, and BP-6, corresponding
to $m_{\tilde{b}_1} = 1000$~GeV - along with the dominant SM
backgrounds. Following these distributions, we can discuss the
optimization of our selection cuts in order to improve the signal to
background ratio.


\subsection{Cut-based Analysis}  

Our cut optimization prescription resembles the one adopted by the CMS 
collaboration \cite{CMS:2014yxa} in order to distinguish a $t\bar{t}b\bar{b}$ sample 
from a background sample of $t\bar{t} +$ jets (upto 2). 
In this section, we consider only leptonic events, i.e. events which have at least
one lepton. In order to distinguish the signal from background,
we make use of five discriminating variables: the number of jets, the $p_T$ of the 
hardest and the second hardest jets in each event, the scalar
$H_T$ and the transverse mass variable, $M_{T2}$. The individual 
variables and the cuts imposed on them are discussed below:

\begin{itemize}
\item ~{\bf C1} : We demand that each signal event contain at least
  one lepton. This particular choice of signal topology will
  substantially remove the most severe SM background coming from the 
  pure QCD multijet processes.

\item ~{\bf C2} : In any process with multiple jets in the final
  state, the number of jets (including both the b-jets and the
    light jets) plays a very crucial role as a discriminatory
  variable. Due to the large mass-splitting between the sbottom and the top, both
    the latter and the other daughter would, often, carry a large
    $p_T$, and, hence, we expect a higher multiplicity of 
   large--$p_T$ jets as compared to the SM backgrounds which, 
   typically, have a significant number of softer jet-progenitors. Consequently, we
   demand that the number of
  jets be greater than four.

\item ~{\bf C3} : Since we expect the non b-tagged jets coming from
  the sbottom decay to have a high $p_T$, we can place a $p_T$ cut on
  such a jet.  We demand that the leading non~b-tagged jet have a $p_T
  > 250$ GeV.

\item ~{\bf C4} : Since the sbottom pair decay produces two light
  jets, we expect that the second hardest non~b-tagged jet will also
  be very energetic.  We put a cut of $p_T > 150$ GeV on the
  sub-leading non~b-tagged jet.  The light jets, if any, from the
  background processes are not expected to have such a high $p_T$.

\item ~{\bf C5} : We calculate $H_T$, the scalar sum of $p_T$ of all the
  visible particles, namely jets, leptons and photons. 
  It is
  defined as follows:
\begin{equation}
H_T = \sum_{i=e,\mu, j, \gamma} \left|\vec{p}_T(i)\right|.
\label{HT_defn}
\end{equation}
The importance of this variable as a signal discriminator is very well
reflected in Fig ~\ref{fig:HTMt2}.  If we demand that our signal
events should have substantially large value of $H_T \sim 1000 $ GeV
then most of the $t{\bar t} jj $ and $t {\bar t} b {\bar b}$ events
are removed. This is again taking advantage of the fact that the large
mass of the sbottom results in jets and leptons with a $p_T$ typically
much higher than those emerging from SM processes.

\item ~{\bf C6} : Finally, we put a cut on the transverse mass
  variable $M_{T2} > 360$ GeV.  The variable is defined as
  \cite{Barr:2003rg}:
\begin{equation}
M_{T2}\left(\vec{p}_T^{V1},\vec{p}_T^{V2},\PET\right) 
= \min_{\slashed{\vec{p}}_T^1+\slashed{\vec{p}}_T^2 = 
\slashed{\vec{p}}_T} \left[\max \left\lbrace M_{T} 
\left(\vec{p}_T^{V1},\slashed{\vec{p}}_T^1\right),
M_{T} \left(\vec{p}_T^{V2},\slashed{\vec{p}}_T^2\right)\right\rbrace \right] 
\end{equation}
where, $\slashed{\vec{p}}_T^1$ and $\slashed{\vec{p}}_T^2$ are two
hypothetical subdivisions of the total missing transverse momentum
$\PET$. The separation of the visible particles into two
  sets with associated transverse momenta $\vec{p}_T^{V1}$ and
  $\vec{p}_T^{V2}$, is done so that the invariant masses of the two
  parts are as close to each other as possible.

In general, the transverse mass 
$M_T \left(\vec{p}_1, \slashed{\vec{p}}_2\right)$ of the 
$\left(\vec{p}_1, \slashed{\vec{p}}_2\right)$ system is defined as
\begin{equation}
M_T \left(\vec{p}_1, \slashed{\vec{p}}_2\right) = \sqrt{m_1^2 + 2 \left| \vec{p}_1\right| \left|\slashed{\vec{p}}_2\right| (1-\rm{cos}\phi)} \ .
\end{equation} 
Here, $\phi$ is the azimuthal angle between the
$\vec{p}_1$ and $\slashed{\vec{p}}_2$ vectors with $\slashed{\vec{p}}_2$ corresponding 
to a massless particle (neutrino) and $m_1^2 \equiv p_1^2$. For the process
under consideration, the visible part comprises of a b-quark, a light
quark and a lepton coming from each of the bottom squarks. Given the
symmetry of the system, we group the visible entities such that the
two visible parts are nearly identical in invariant mass.  For
calculating $M_{T2}$, we use the Cheng and Han Bisection algorithm
\cite{Cheng:2008hk}. From the distribution shown in
Fig.~\ref{fig:HTMt2}, we can easily see that this variable too has a
good discriminatory power.
\end{itemize}
%

The event summary for the signal and backgrounds after individual
selection cuts is presented in Table~\ref{cut-flow}.  The numbers in
the table denote the resulting cross-sections after each selection cut
is applied to both signal and background events. The first row in the
table, denoted by `C0', refers to the NLO production cross-section for
each process.

The numbers on the subsequent rows
relate to the surviving cross-section for each of the cases after the
relevant cut (indicated as bullet points earlier) has been imposed.

\begin{table}[htb!]
 \centering
\begin{tabular}{|c|ccccc|cccccc|}
 \hline 
 Cuts & $t\bar{t} + jets$ & $t\bar{t}b\bar{b}$ & $t\bar{t}Z$ & $t\bar{t}H$ & $t\bar{t}W$ & BP1 & BP2 & BP3 & BP4 & BP5 & BP6 \\
 \hline \hline 
 C0: & $8.3  \times 10^5$ & $1.7  \times 10^4$ & $8.7  \times 10^2$  & $5.1  \times 10^2$  & $6.5  \times 10^2$  &  $5.2  \times 10^2$  & $1.8  \times 10^2$  & 67.0 & 28.3 & 12.9 & 6.2  \rule{0pt}{2.6ex} \\[1.5mm]
 \hline  
 C1: & $1.8 \times 10^5$ & $3.3 \times 10^3$ & $2.7  \times 10^2$ & $1.0  \times 10^2$ & $2.3  \times 10^2$ &  90.6 & 29.6 & 10.9 & 4.4  & 1.9  & 0.8 \rule{0pt}{2.6ex}\\[2mm]
 C2: & $3.8 \times 10^4$  & $1.2 \times 10^3$   & $1.4  \times 10^2$  &  63.4 &   89.4 &  76.8 &  25.8 & 9.7  & 3.9  & 1.7  & 0.8  \\[2mm]
 C3: & $3.9 \times 10^3$ & 65.2    & 20.3  &   6.8  &   10.5 &   43.3 &  19.3 & 8.2  & 3.6  & 1.6  & 0.7  \\[2mm]
 C4: & $1.6 \times 10^3$  & 27.2    & 11.0  &   3.2  &   2.1  &   33.4 &  16.1 & 7.2  & 3.2  & 1.5  & 0.6  \\[2mm]
 C5: & $9.6 \times 10^2$   & 16.3    & 7.7  &   2.1  &   3.1  &   26.1 &  14.3 & 6.9  &  3.2  & 1.4  & 0.6  \\[2mm]
 C6: & $7.6 \times 10^2$   & 13.9    & 5.5   &   1.6  &   1.9  &   17.4 &  10.9 & 5.6  &  2.7  & 1.3  & 0.6  \\[2mm]
 \hline
\end{tabular}
\caption{\small {\it The surviving cross-section (in fb) for
    the different processes after each of the cuts. For $t\bar{t}$ +
    jets, we consider up to 2 jets.}}
\label{cut-flow}
\end{table}

We can now estimate the signal significance corresponding to each
benchmark point at the 13 TeV LHC assuming 300 fb$^{-1}$ of
integrated luminosity. We are interested in the cross-section after the cut `C6'
is imposed (last row of Table \ref{cut-flow}).  The number of signal
(background) events, denoted by $S$ ($B$), is given by the product of
this cross-section and the integrated luminosity.  In Table \ref{significance},
we tabulate the signal significance $\mathcal S$ given by $$\mathcal S
= {S \over \sqrt{S+B}} \ .$$
 
\begin{table}[htb!]
\centering
\begin{tabular}{|c|ccccccc|}
\hline
 & Background & BP1 & BP2 & BP3 & BP4 & BP5 & BP6 \rule{0pt}{2.0ex}\\
\hline
$m_{\tilde{b}_1}$(GeV) &  & 500 & 600 & 700 & 800 & 900 & 1000\\
\hline
$\mathcal{N}$ & $2.3  \times 10^5$ & $5.2  \times 10^3$ & $3.3  \times 10^3$ & $1.7  \times 10^3$ & $8.2  \times 10^2$ & $3.9  \times 10^2$ & $1.7 \times 10^2$ \rule{0pt}{2.6ex}\\
$\mathcal{S} = \frac{S}{\sqrt{S+B}}$& & 10.7  & 6.7  & 3.5  & 1.7  & 0.8  & 0.4  \\
\hline
\end{tabular}
\caption{\small {\it Number of background and signal events 
for a integrated luminosity of $300 ~{\rm  fb^{-1}}$, 
along with the significances 
for the different benchmark points. See text for details.}}
\label{significance}
\end{table}

It is evident from the table that, for an integrated
luminosity of 300 ${\rm fb^{-1}}$, the LHC stands in extremely good
stead to detect the sbottom should its mass be 600 GeV or below. The
LHC will graze past the exclusion limit of 95\% C.L. for masses around
$\sim 750$~GeV. Given 3000 ${\rm fb^{-1}}$ integrated luminosity, we find that
the discovery reach (i.e. $5\sigma$ significance) will exceed 800 GeV
and the exclusion bounds might be extended to beyond the 900 GeV mass
point.

\subsection{Multivariate Analysis}

To achieve a better discrimination between the signal and the SM
background, we perform a multivariate analysis (MVA)
using the Boosted Decision Tree (BDT) algorithm as implemented in the
Toolkit for Multivariate Data Analysis (TMVA) \cite{Hocker:2007ht}
with ROOT \cite{Antcheva:2009zz}. We briefly describe the procedure,
the details of which may be found in Ref. \cite{Hocker:2007ht}, along with
the parameters for our analysis below. 

Decision trees are used to classify events as either signal-like or
background-like. Each node in a decision tree uses a single
discriminating variable, along with a certain cut value imposed on it,
to provisionally classify events as either signal-like or
background-like depending on the purity of the sample. The decision
tree needs to be `trained' and that starts with the root node. We can
think of the process as two bins originating from the root node (i.e.,
the zeroth node), one having events classified as signal-like and the
other as background-like. At the next level, each of these bins can be
treated in exactly the same way as the root node, using a variable of
choice and a particular value of cut on it, giving us two bins---one
signal-like and the other background-like---for each node.  A tree is
built up to a depth either determined by the remaining number of
background events, or by the depth specified by the user. The final
leaf nodes contain background-like and signal-like events from the
training sample. Generally half of the provided sample is used for
training and the other half is then used for testing.

Decision trees, however, are unstable under statistical fluctuations
and cannot be used as good classifiers. Instead, the technique of
boosting can be used to combine several classifiers into a single one,
such that the latter is more is stable under such fluctuations and,
hence, has a smaller error than the individual ones. Boosting modifies
the weights of individual events and creates a new decision
tree. Higher weights are preferentially assigned to the incorrectly
classified events. Previously assigned weights are modified by
$\alpha$, given by
\begin{equation}
\alpha = \frac{1 - \epsilon}{\epsilon}, \quad \quad {\rm where } \quad \epsilon = \sqrt{\frac{p(1-p)}{N}} \ ,
\end{equation}
where $N$ is the total number of training events in the node and $p =
S/(S+B)$, called the purity of the sample. The number of decision
trees in the forest we use is given by {\cmssfont NTrees =} 400, the
maximum depth of the decision tree allowed is {\tt MaxDepth =} 5
and the minimum percentage of training events in each leaf node is given by 
{\tt MinNodeSize =} 2.5\%.  We choose Adaptive Boost, proven
to be effective with weak classifiers and implemented as {\tt
  AdaBoost} in TMVA, as the method for boosting the decision trees in
the forest with the boost parameter $\beta \equiv \mbox{\tt
  AdaBoostBeta} = 0.5$. This parameter adjusts the learning rate of
the algorithm simply by changing the weights $\alpha \to
\alpha^\beta$. We have used the default values of the BDT parameters, viz. 
{\tt NTrees}, {\tt MaxDepth} and {\tt MinNodeSize}.


\begin{table}
\centering
{\large
\begin{tabular}{ |l|l| }
\multicolumn{2}{l}{} \\
\hline
\bf{Set-1} & $(p_T)_{j1}$, $(p_T)_{j2}$, $(p_T)_{j3}$, $(p_T)_{j4}$, $(p_T)_{bj1}$, $(p_T)_{bj2}$, $(p_T)_{bj3}$, $(p_T)_{bj4}$, \\[1mm]
 & $H_T$, $\MET$, nJets, nbJets, $M_{T2}$, $m^{h}_{\tilde{b}_1}$, $m^{\ell}_{\tilde{b}_1}$ \\ \hline
\bf{Set-2} & $(p_T)_{j1}$, $(p_T)_{j2}$, $(p_T)_{j3}$, $(p_T)_{j4}$, $(p_T)_{bj1}$, $(p_T)_{bj2}$, $H_T$,\\[1mm]
 &  $\MET$, nJets, $M_{T2}$, $m^{h}_{\tilde{b}_1}$, $m^{\ell}_{\tilde{b}_1}$  \\ \hline
\bf{Set-3} & $(p_T)_{j1}$, $(p_T)_{j2}$, $(p_T)_{bj1}$, $(p_T)_{bj2}$, $H_T$, $\MET$, nJets, $M_{T2}$ \\[1mm] \hline
\bf{Set-4} & $(p_T)_{j1}$, $(p_T)_{j2}$, $H_T$, $\MET$, nJets, $M_{T2}$ \\
\hline
\end{tabular}}
\caption{{\small{\it Different sets of variables that can be considered for the multivariate analysis. We choose Set-2 for our analysis.}}}
\label{tab:vars}
\end{table}

\begin{figure} [!htb]
\includegraphics[scale=0.52]{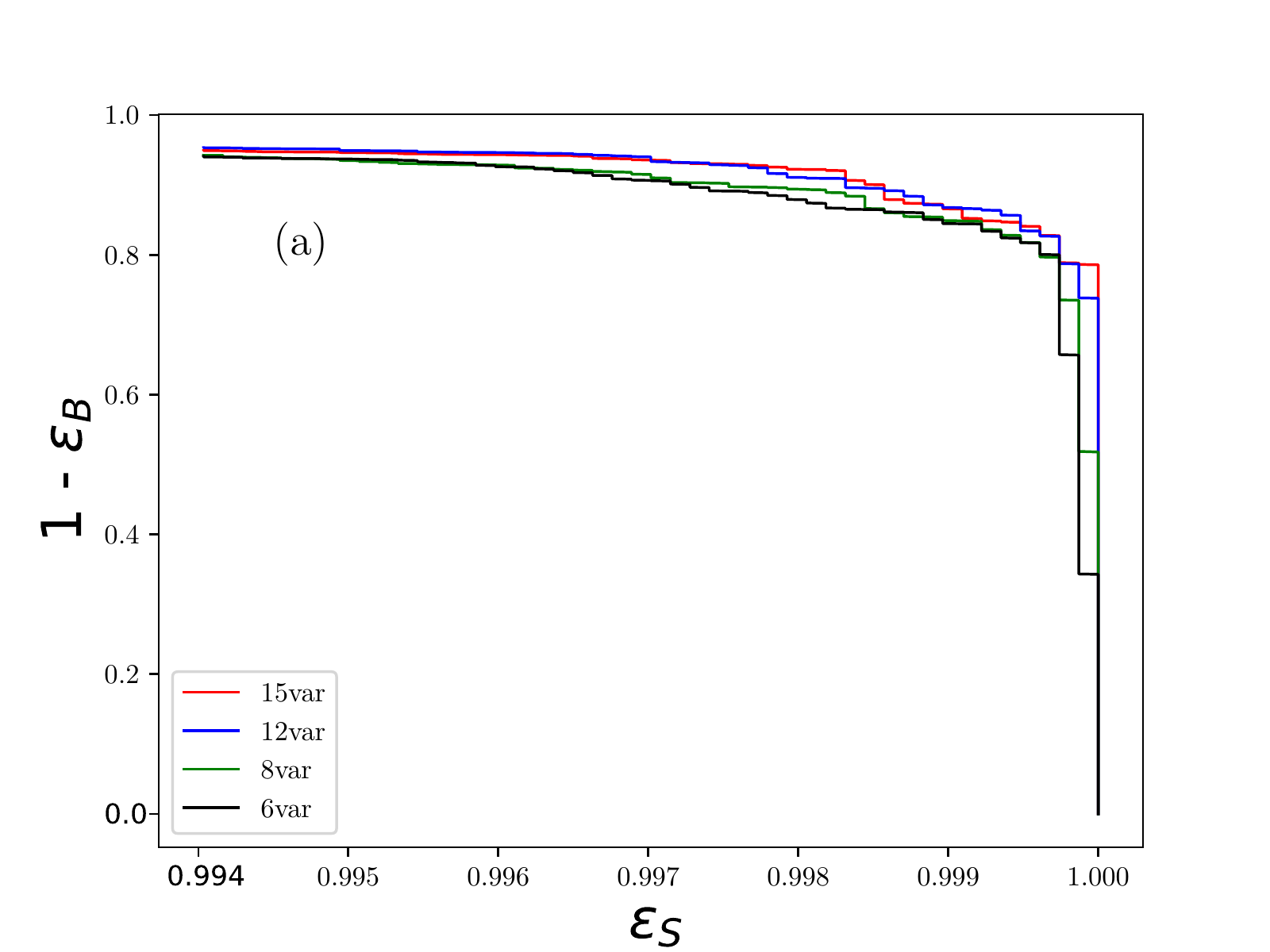}
\includegraphics[scale=0.52]{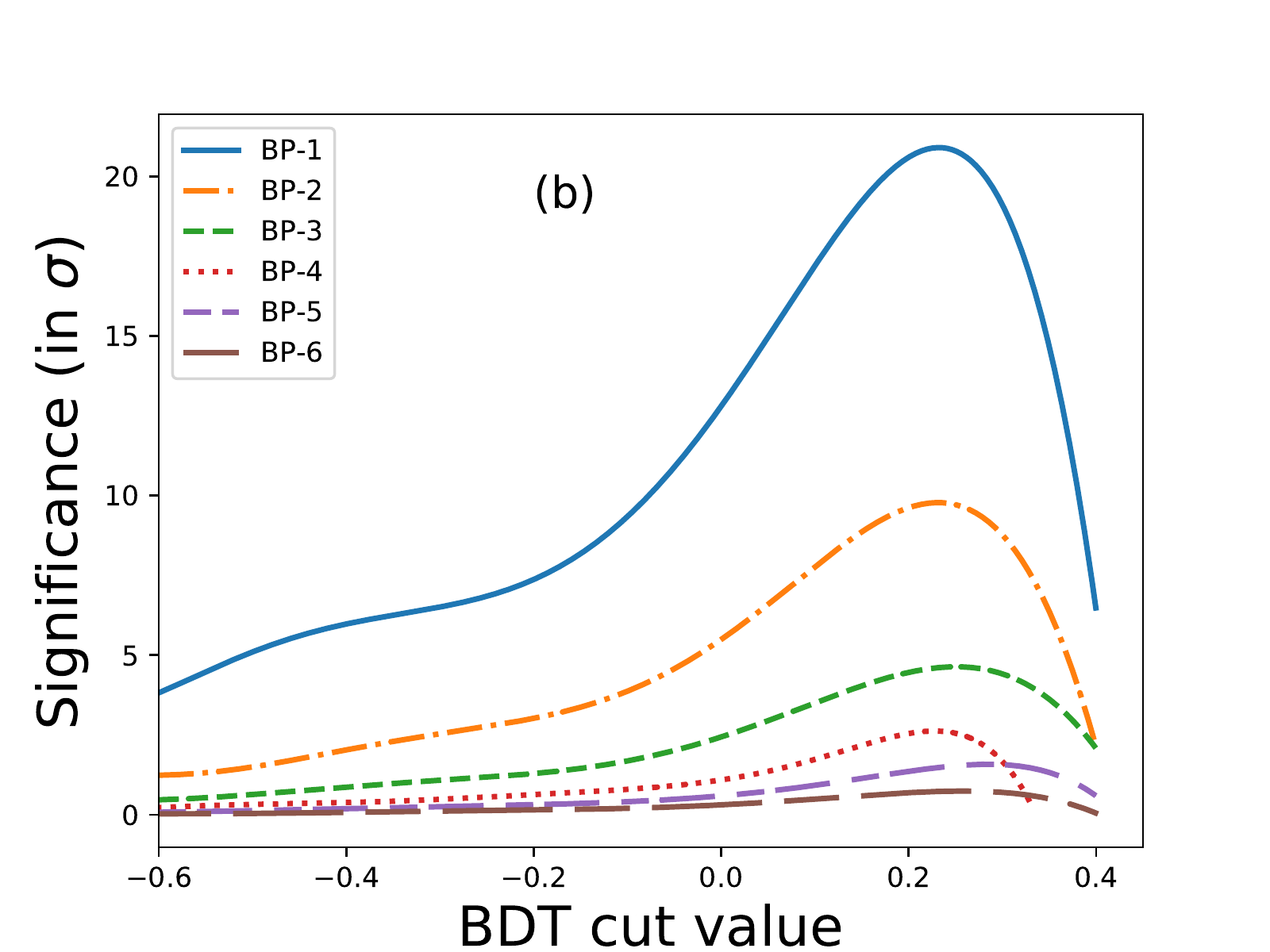}
\caption{\small {\it On the left ($a$), the ROC plot is shown for BP-4 with the four sets of variables and 
on the right ($b$), the plot for the signal significance against different BDT cut-values is shown.}}
\label{roc_significance}
\end{figure}

A challenge endemic to TMVA is finding
  an optimal set of observables that would lead to the best possible 
  discrimination between signal and background events.  It is important to note that
  a larger set of variables need not always provide better
  discrimination, especially if it is mostly filled with irrelevant
  observables.  We tried four sets comprising of 15 (Set-1), 12
  (Set-2), 8 (Set-3) and 6 (Set-4) variables respectively as detailed
  in Table \ref{tab:vars}. We then plot the ROC (Receiver's Operative
  Characteristic) Curve for these sets. The ROC curves signify the
  efficiency of the signal ($\epsilon_S$) with respect to the
  efficiency of rejecting the background ($1-\epsilon_B$), with
  $\epsilon_B$ being the efficiency of the background. This is
    exemplified by the left panel of Fig.\ref{roc_significance}, wherein we plot these ROCs
  for the benchmark point BP-4. 
  Whereas the use of Set-1 and Set-2 offers some improvement
    over Sets 3 and 4, the former are virtually indistinguishable in
    their efficacy. In other words,
    the extra variables in Set-1 are of very little relevance. Given
    this, we choose the largest set of variables without keeping any
    irrelevant variables, namely Set-2 for the rest of the analysis in
    this section.

The variables chosen as BDT inputs have already been introduced in the
previous section (see the cut-based analysis) except for the two new
variables, namely $m^{h}_{\tilde{b}_1}$ and $m^{\ell}_{\tilde{b}_1}$, which represent 
the reconstructed sbottom mass using the hadronically ($h$) and 
leptonically ($\ell$) decaying top quarks 
respectively. We select events with exactly one
isolated lepton (electron or muon) with two or more b-tagged jets,
utilising only the two hardest b-tagged jets in our
reconstruction. Additionally, we work with the four hardest light  ({\em i.e.}, non-b-tagged) jets in the event.

We could have also attempted to reconstruct 
the sbottom for events with two isolated leptons originating from the 
leptonic decay of the two top quarks. 
However, the presence of two neutrinos, the only source of missing energy 
here, renders the reconstruction 
non-trivial and makes it a highly involved task. 
 With the dileptonic branching
  fraction being only 5\% (compared to 30\% for the semileptonic one),
  and with the pair production cross-section falling rapidly with the
  sbottom mass, this channel is likely to be important only in the
very high luminosity run of the LHC.  In this work, we thus focus only
the semi-leptonic case when events contain exactly one isolated lepton
with b-tagged and light jets and missing transverse energy.  The
interested reader can refer to \cite{Simak:1999gka,Bai:2008sk} for the
detailed implementation of the reconstruction of $t\bar{t}$ and heavy
resonances using dileptonic modes.

Before we proceed to reconstruct the top quark, we must
reconstruct the $W$ bosons. The hadronically decaying $W$ boson is 
reconstructed by choosing the pair of light non b-tagged jets which 
give an invariant mass closest to the actual $W$ boson mass with a 
further demand that the thus reconstructed
mass lies within $M_{W} 
\pm 30$~GeV. 
The leptonically decaying $W$ boson in the decay of the top quark is 
reconstructed, within a quadratic ambiguity, 
from the four momentum of the lepton, $p_\ell$ and the
missing transverse momentum $\vec {\cancel {\rm p}}_T \equiv (\PEx, \PEy)$, 
by imposing 
the condition that the invariant mass $M_{\ell \nu} = M_W$. Note that, here it 
is assumed that the only source of missing energy is the neutrino originating 
from the leptonic decay of W. Using the 4-vector of the isolated lepton $p^{\mu} = (E^{\ell},
p_x^\ell, p_y^\ell, p_z^\ell)$, arising from the decay of the $W$, one
can construct the longitudinal component (and, hence, the energy)
 of the missing momentum as follows:
\begin{equation}
\PEz = \frac{1}{2 \left(E^{\ell 2} - p_z^{\ell 2}\right)} \left[p_z^\ell
\left(2 p_x^\ell \PEx + 2 p_y^\ell \PEy - m_\ell^2 + M_W^2\right)
\pm \sqrt{\Delta}\right],
\label{eqn:pzMET}
\end{equation}
where the quantity $\Delta$ is given by
\begin{equation}
\Delta = E^{\ell 2} \left[\left(2 p_x^\ell \PEx + 2 p_y^\ell \PEy - m_\ell^2 + M_W^2\right)^2 
- 4 \PET^2 \left(E^{\ell 2} - p_z^{\ell 2}\right)\right], 
\end{equation}
with $m_\ell$ being the mass of the lepton and $M_W$ being the input
mass for the $W$ boson. This provides us with two values of 
$\PEz$ corresponding to the two signs of the square root.
For certain configurations, however, one
may obtain $\Delta < 0$, rendering the calculated $\PEz$
complex and thus unphysical. In these cases, one can re-calculate the
missing energy by finding those values of $\PET$ for which $\Delta \ge
0$:
\begin{equation}
\PET = \frac{1}{2\left(E^{\ell 2} - p_z^{\ell 2} - (p_x^\ell \cos \phi + 
p_y^\ell \sin \phi)^2\right)} \left[-(p_x^\ell \cos \phi + p_y^\ell \sin \phi)
(m_\ell^2 - M_W^2) \pm \sqrt{(m_\ell^2 - M_W^2)^2 (E^{\ell 2} - p_z^{\ell 2})} \right].
\label{eqn:pt_negDelta}
\end{equation}

For each sign of the square root in Eqn. \ref{eqn:pt_negDelta}, we get 
a value of $\PET$, which when substituted in Eqn. \ref{eqn:pzMET} give
two values of $\PEz$ for every value of $\PET$. Thus, we end up with
four values of $\PEz$ in this case, instead of just two as in the 
earlier case. 

For each value of the z-component of the MET (i.e. $\PEz$), we can
reconstruct the leptonically decaying top quark mass by combining the
4-momenta of the lepton, b-jets and the missing energy. Several 
reconstructed mass combinations can exist, depending on the number
of solutions of $\PEz$ and since there are two b-tagged jets to 
choose from.

To obtain the optimal
values of the leptonic and hadronic top quark masses in each event, a
minimum-$\chi^2$ approach is adopted with the $\chi^2$ defined as:
\begin{equation}
\chi^2 = \frac{{\left(m_{tH} - m_t\right)}^2}{\sigma^2_{m_{tH}}} + \frac{{\left(m_{tL} - m_t\right)}^2}{\sigma^2_{m_{tL}}},
\end{equation}
where $\sigma_{m_{tL}}$ and $\sigma_{m_{tH}}$ represent the
uncertainty in top quark mass measurement for leptonically and
hadronically decaying tops respectively at the LHC.  We consider
$\sigma_{m_{tL}} = 2.7$~GeV and $\sigma_{m_{tH}} = 1.15$~GeV
\cite{Aaboud:2017mae,Sirunyan:2017uhy}. Using the 4-momentum
information of the isolated lepton and missing transverse energy along
with the two b-tagged jets and the leading four non b-tagged jets, we
reconstruct the leptonic and hadronic top quark masses. The
combination which leads to the lower $\chi^2$ value is chosen.
%
Nice resonance peaks around top quark mass are observed for both the leptonically 
and hadronically decaying tops for all the benchmark points. 

\begin{figure} [!htb]
\includegraphics[scale=0.5]{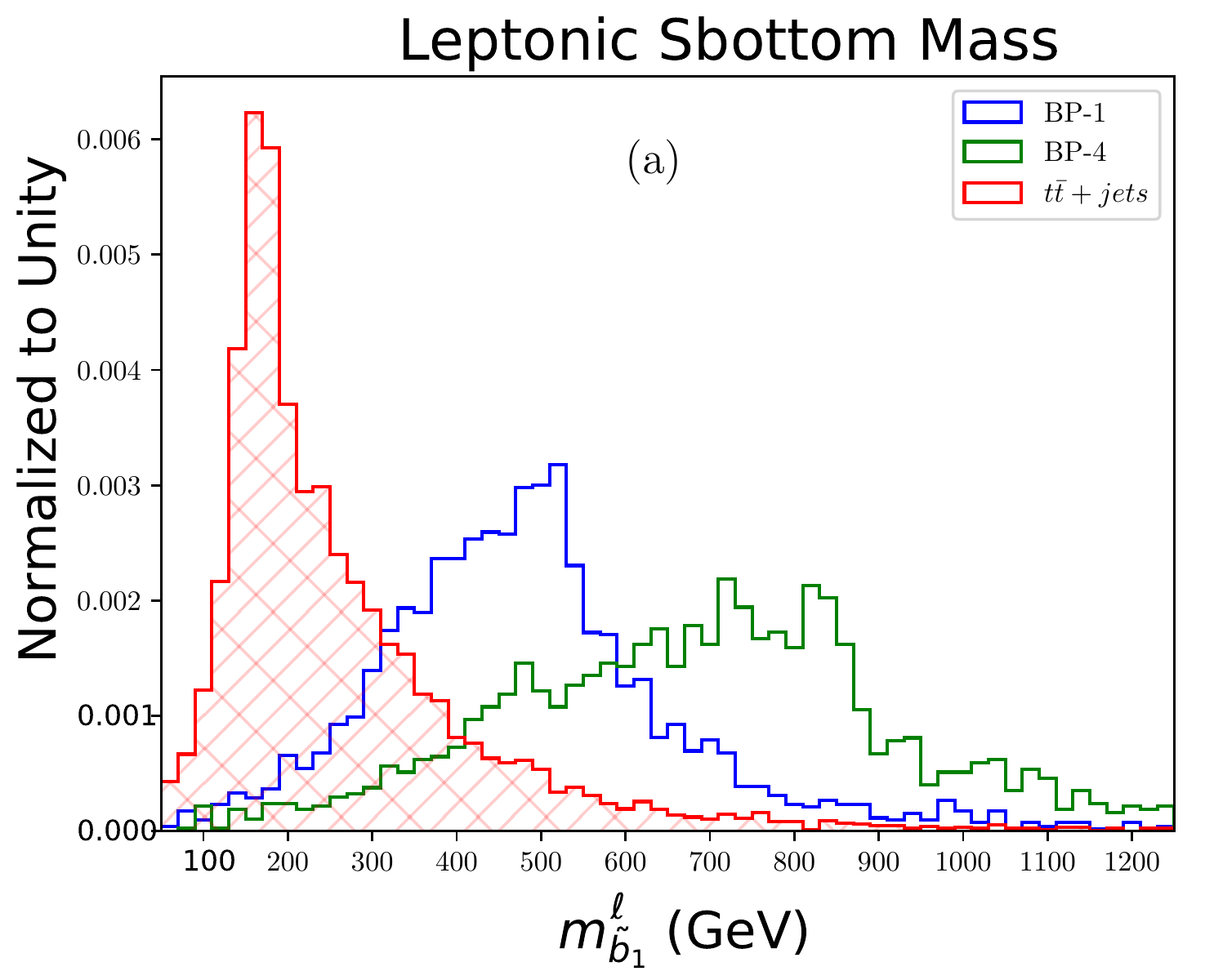}
\includegraphics[scale=0.5]{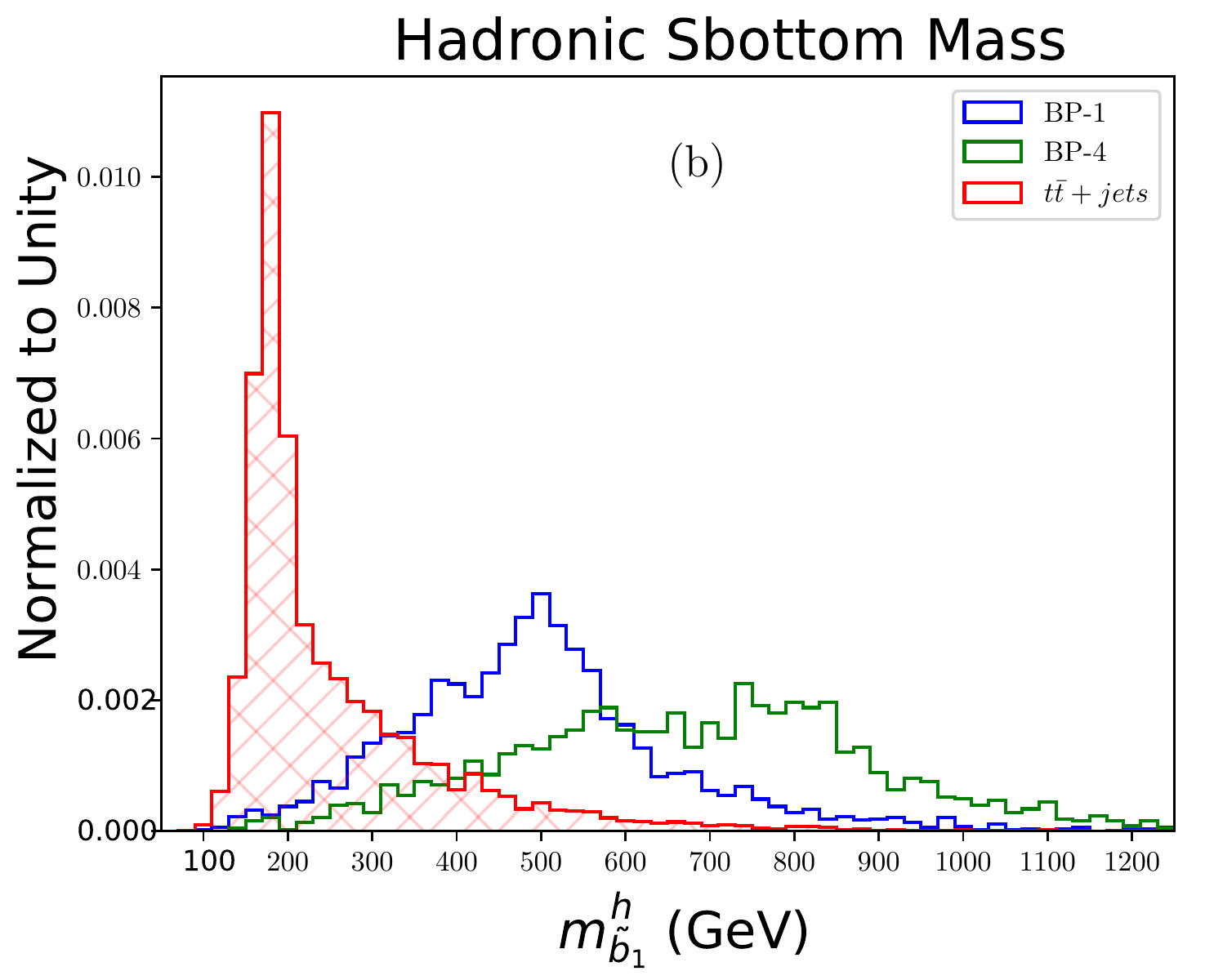}
\caption{\small {\it Reconstructed sbottom masses for BP-1, BP-4 and the $t\bar{t}$ events, 
left ($a$) denotes the case when the top quark decays leptonically while the right ($b$) 
signifies the hadronically decaying top quark scenario. For details, see the text.}}
\label{fig:sbot_reco}
\end{figure}

After the reconstruction of two top quarks, we are now left with the 
final reconstruction of the sbottom mass using these two reconstructed top quarks and 
and the two remaining light quark jets originating from the decay of the two 
sbottoms. 
For each reconstructed top mass, there are two possible choice to combine
the light jets for the reconstruction of the sbottom mass.
We select the combination which 
leads to the least difference 
between the reconstructed mass of the leptonically decaying sbottom
and the hadronically decaying sbottom.  The plot for the reconstructed
sbottom for BP-1 (corresponding to a 500 GeV sbottom) and for BP-4
(corresponding to a 800 GeV sbottom) are shown in
Fig. \ref{fig:sbot_reco}, where the left and right panels denote the
reconstruction method involving the leptonically and hadronically
decaying top quarks. The reconstructed sbottom masses peak at the
truth masses for the two benchmark points, while for $t\bar{t}$ events
it peaks near the truth top quark mass. The peaks corresponding to the
signal events are significantly distinct from that of the backgrounds,
and this motivates us to consider the reconstructed masses as the BDT
inputs.

\begin{table}[htb!]
\centering
\begin{tabular}{|l|cccccc|}
\hline
& BP-1 & BP-2 & BP-3 & BP-4 & BP-5 & BP-6\\
\hline
\hline 
$m_{\tilde{b}_1}$(GeV) & 500 & 600 & 700 & 800 & 900 & 1000  \\[1mm]
\hline
BDT cut\quad  \quad & 0.231 \quad  & 0.234 \quad  & 0.258 \quad & 0.230 \quad  & 0.311 \quad  & 0.294 \\
$\mathcal{S} = \frac{S}{\sqrt{S+B}}$ & 20.9 & 9.9 & 4.7 & 2.7 & 1.6 & 0.9 \\
\hline
\end{tabular}
\caption{\small {\it Signal significances for 
the benchmark points with the choice of BDT cuts with 
$\mathcal L = 300~{\rm fb^{-1}}$ of integrated luminosity.}}
\label{tab:bdt2}
\end{table}

\begin{figure}
\centering
\includegraphics[scale=0.55]{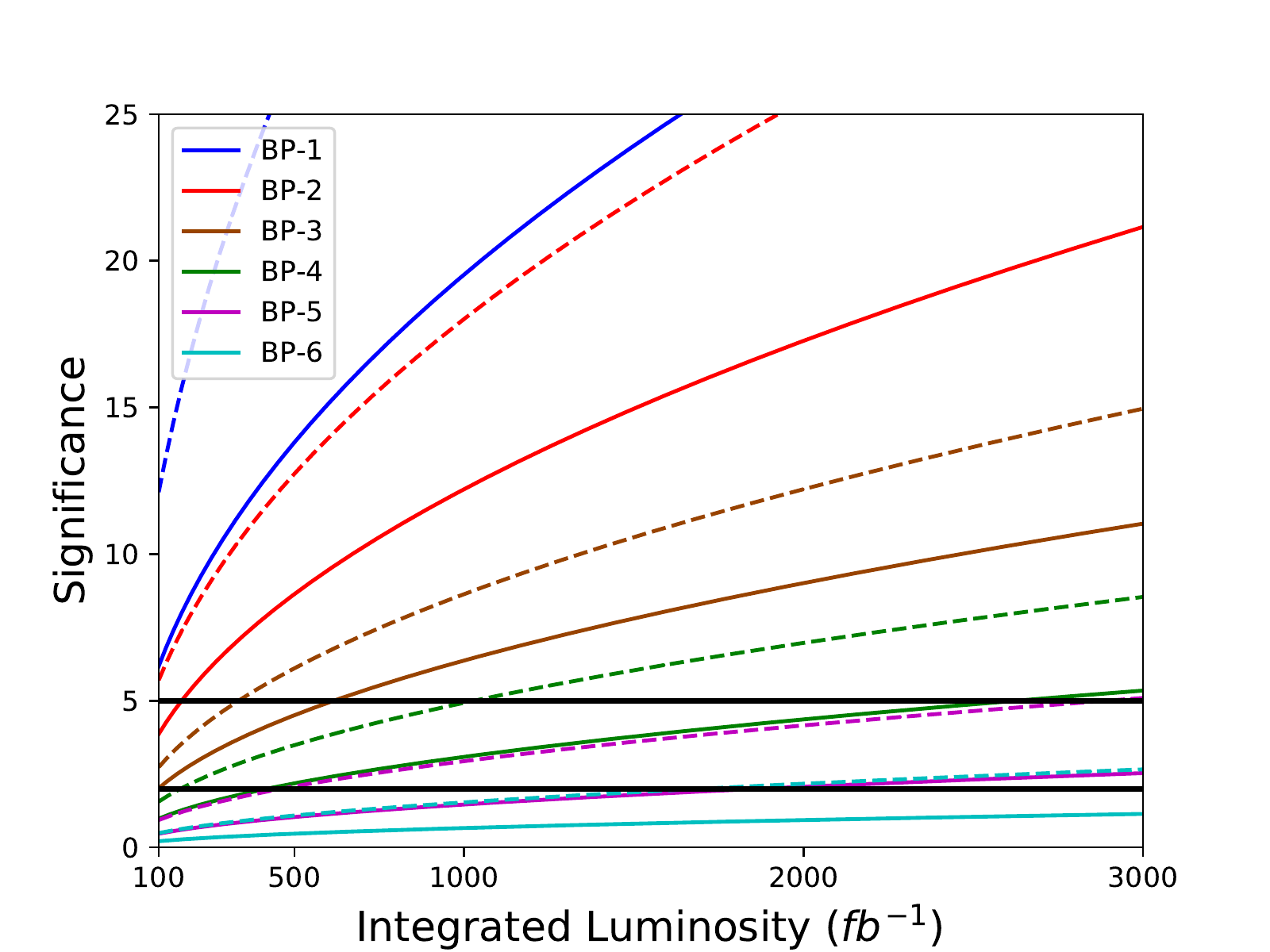}
\caption{{\small{\it Plot of significance versus the integrated luminosity. While
      the inset legend shows the colour for the different benchmark
      points, the solid line corresponds to the significance
      corresponding to the cut-based analysis and the dashed line to
      that provided by the multivariate analysis. Horizontal lines at
      $2\sigma$ and $5\sigma$ indicate the potential for exclusion and
      discovery.}}}
\label{fig:siglum}
\end{figure}

For each benchmark point, the variation of signal significance with the 
BDT cut value has been shown in plot (b) of Fig.~\ref{roc_significance} and Table \ref{tab:bdt2}
shows the best signal significance with the corresponding BDT cut values assuming
$\mathcal{L} = 300{\rm\ fb}^{-1}$. Clearly, the MVA improves the reach of 
the search compared to the cut-based analysis, e.g. signal significance
improves from 10.7 to 20.9 for BP-1, resulting in an increase of the discovery reach 
of $\sim 100 {\rm \ GeV}$ in the mass of the $\tilde{b}_1$. 
The variation of signal significance with integrated luminosity is shown in 
Fig.~\ref{fig:siglum}, solid lines for the cut-based analysis and the dashed lines 
for the MVA. One can observe that with 2000 ${\rm fb}^{-1}$ data, $\tilde{b}_1$ 
mass of upto 1 TeV can be explored at the High Luminosity run of LHC.

\section{The Hadronic  Final State} 
We now consider the case where both the top-quarks decay fully
hadronically. The fully hadronic final state is difficult to investigate
at the LHC because of the overwhelming QCD background. However, in our
signal events, the top quarks are expected to be boosted, such that the
three quarks from its decay form a `fatjet' with substructure. Our plan
is to exploit the substructure of such a fatjet to identify a top quark
and investigate the reach for sbottom using 13 TeV data from the LHC.

Our final state will contain
only reclustered fatjets and we shall attempt to tag some of these
jets as tops. The background should ideally have contributions from
all the SM processes we considered in the leptonic 
counterpart---$t\bar t$ + jets (upto two), $t \bar{t} b \bar{b} $,
$t \bar{t} Z$, $t \bar{t} W$ and $t \bar{t} H$---in addition to the 
QCD
multijet, but for all practical purposes the QCD multijet processes
and $t\bar t$ +jets (upto two) contribute so 
overwhelmingly to the
background (even after cuts) that we really need not consider the other
processes. In this section, we work with this 
simplifying assumption about the background. It is to be noted
    that while simulating the QCD multijet events, we restrict
    ourselves up to four jets at the parton level (light quarks and
    gluons only) due to our computational limitations. However, 
    once parton showering is switched on, the jet multiplicity 
    can and does become larger.

    Our strategy is to tag at least one top quark in each signal event.
     For this purpose, we use {\tt
      HEPTopTagger}\cite{heptoptagger}, which is quite efficient for
    tagging tops with moderate boosts ($p_T \gtrsim 200$~GeV). We
    avail of the energy flow of the particles, provided in the {\tt
      EFlow} branch of the {\tt DELPHES} generated ROOT file to obtain
    the particle information.  We use {\tt FASTJET} to construct fat
    jets of R = 1.8 using the anti-kT jet algorithm with a minimum
    $p_T$ of 30~GeV. The jets with $p_T > 200$~GeV and $|\eta|<3$ are
    then selected to pass through the {\tt HEPTopTagger}.  Before they
    enter the toptagger, these jets are reclustered exclusively with
    the Cambridge-Aachen (C/A) algorithm with the same
    jet radius (viz. R =1.8).  The default settings
    of {\tt HEPTopTagger} were used: the mass drop required for jet
    splitting was set at ${\rm min}(m_{j_1},m_{j_2})/m_j = \mu < 0.8$
    with the minimum mass of a subjet $m_{sub}^{min} = 30$~GeV, where
    $j_1$ and $j_2$ are the subjets of the fatjet $j$.  The top-- and
    W--masses are reconstructed on a set of filtered subjets numbering
    no more than $N_{\rm filt} = 5$. Tops are tagged with masses in
    the range between $m_{top}^{min} = 140$~GeV and $m_{top}^{max} =
    200$~GeV. We achieve an efficiency of about 30\% using these
    conditions for moderate ($\sim$ 200 GeV) to high (say 600 GeV or
    more) $p_T$ regime. The choice of large jet radius indicates that
    we are required to incorporate some jet grooming technique in
    order to get rid of soft and large angle radiations as well as
    underlying events. In our analysis, we use a particular 
technique, named Jet Trimming \cite{Cacciari:2008gd} which has
    been found to be very effective in grooming large R jets. This
    grooming technique involves two independent parameters, namely
    ${\rm R_{trim}}$ and ${p^{\rm frac}_{T}}$. The prescription is to
    essentially recluster the constituents of a given jet with a
    smaller jet radius ${\rm R_{trim}}$ and then keep those subjets
    with $p_T$ greater than a fixed fraction, $p_T^{\rm frac}$ of the
    input jet $p_T$. In our analysis, we optimize these two parameters
    and choose ${\rm R_{trim}} = 0.4$ and ${ p^{\rm frac}_{T}} =
    1\%$. These trimmed jets, obtained after trimming the original
    anti-kT jets, are used for further analysis.

\subsection{Multivariate Analysis}

After passing the jets to the {\tt HEPTopTagger},
we select the events containing at least one top-tagged jet.  
The complete event information is used to construct different
observables, and these, in turn, are used to perform a
multivariate analysis using the TMVA framework. 

Once we have successfully described the full event information in
terms of jets, we classify the different types of jets as top-tagged
jets, b-tagged jets and ``light" jets (non top-tagged, non b-tagged
jets).  For b-tagging, we calculate the angular distance between a jet
and the b-hadron, and make sure that the separation $\Delta R <0.5$.
Furthermore, we also take into account a $p_T$ dependent
b-tagging efficiency given by
\cite{ATLAS_btag}:
\begin{equation}
\epsilon_b = 
\left\{ \begin{array}{ll} 0.5~ & {\rm for} \ \ p_T^b \leq 50~{\rm GeV} \\[0.5ex]
                          0.75 & {\rm for} \ \ 50~{\rm GeV} < p_T^b \leq 400~{\rm GeV} \\[0.5ex]
                          0.5~ & {\rm for} \ \ p_T > 400~{\rm GeV}
               \end{array} \right.  
\end{equation}
Note that the above-mentioned efficiencies are conservative estimates;
with more data and improved algorithms we expect significant
improvement in b-tagging efficiencies. Finally, jets which are not
tagged either as `top-jets' or `b-jets' are called `light jets'.

\begin{figure}[htb!]
\hspace{-6mm}\includegraphics[scale=0.28]{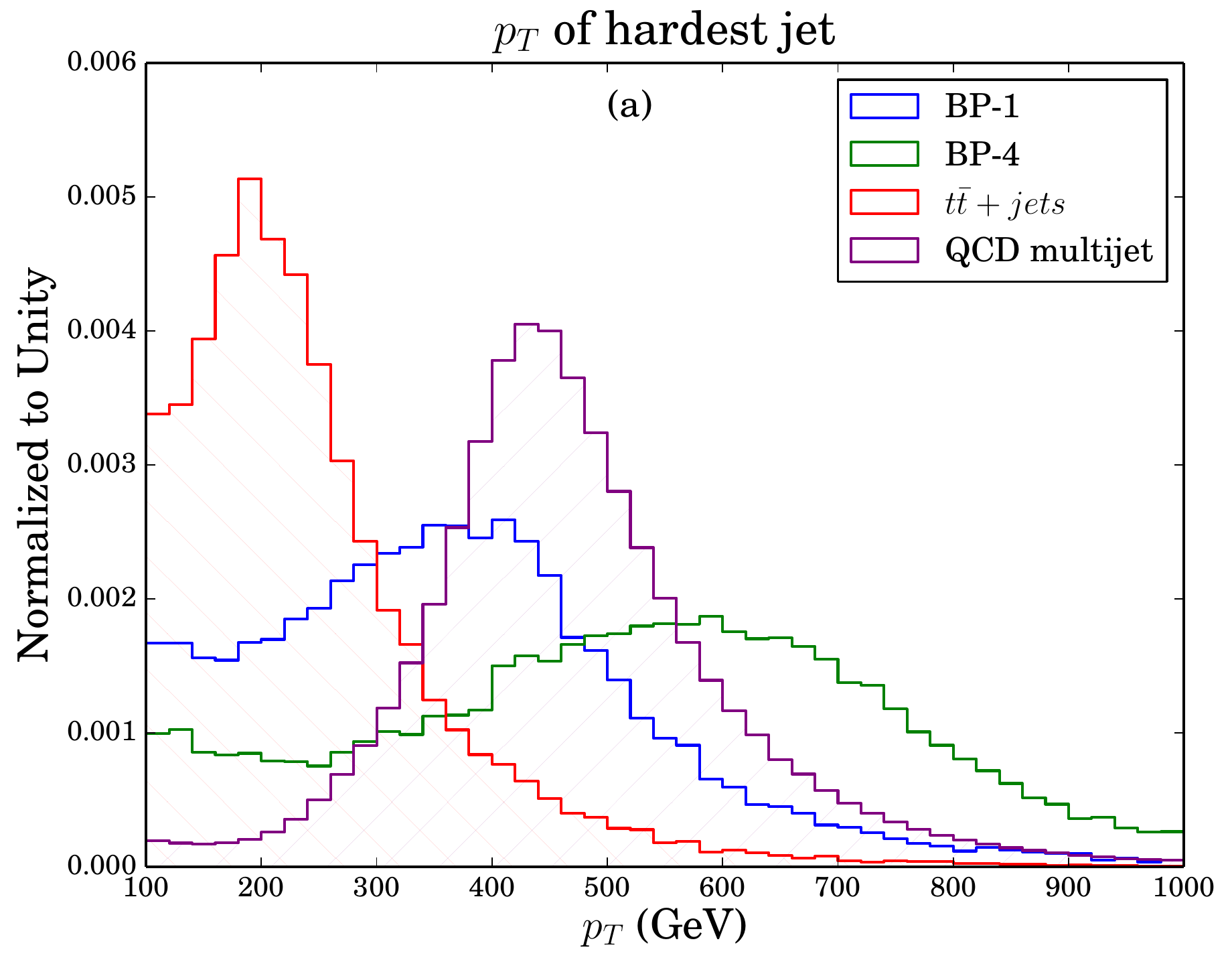}
\includegraphics[scale=0.28]{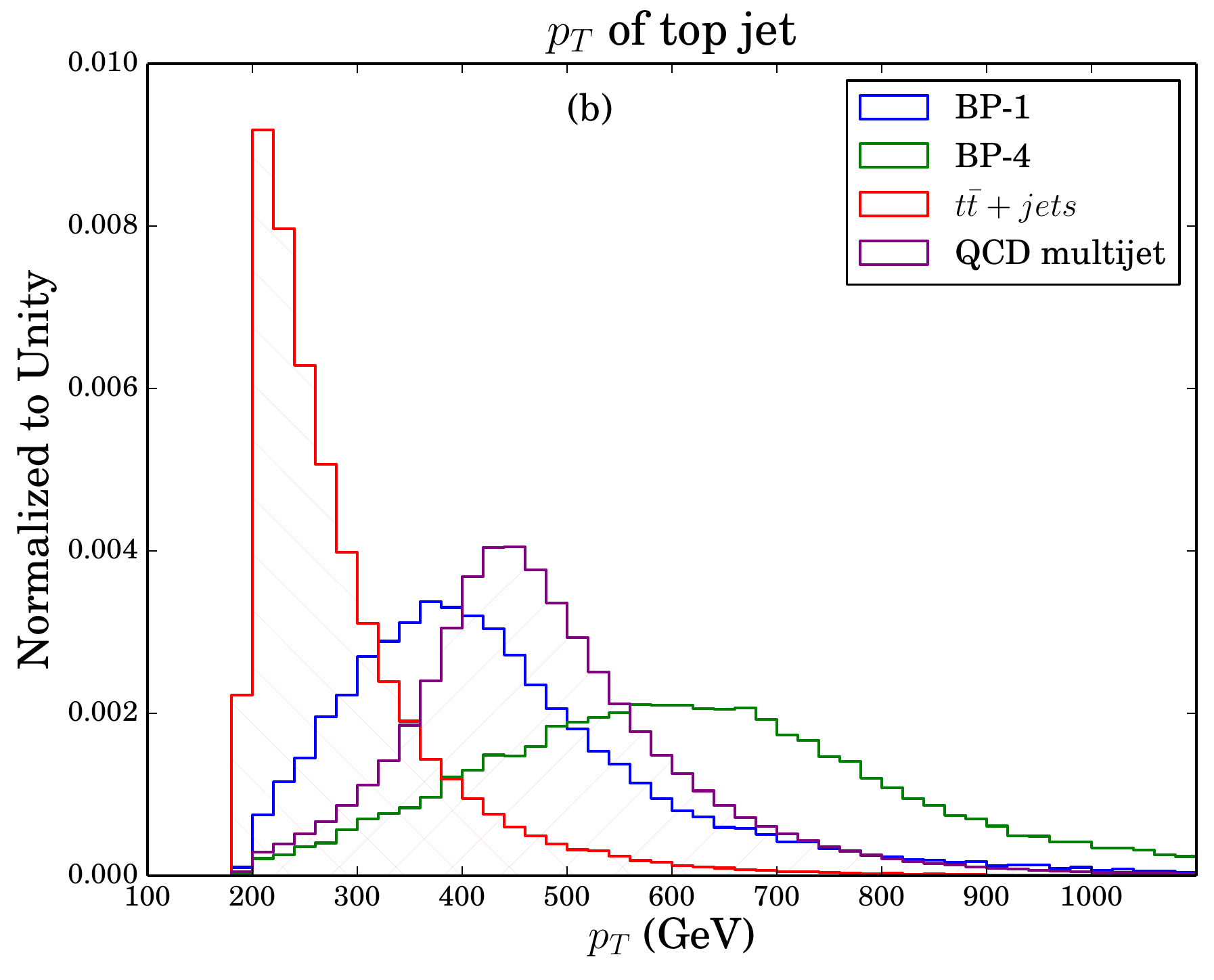}
\includegraphics[scale=0.28]{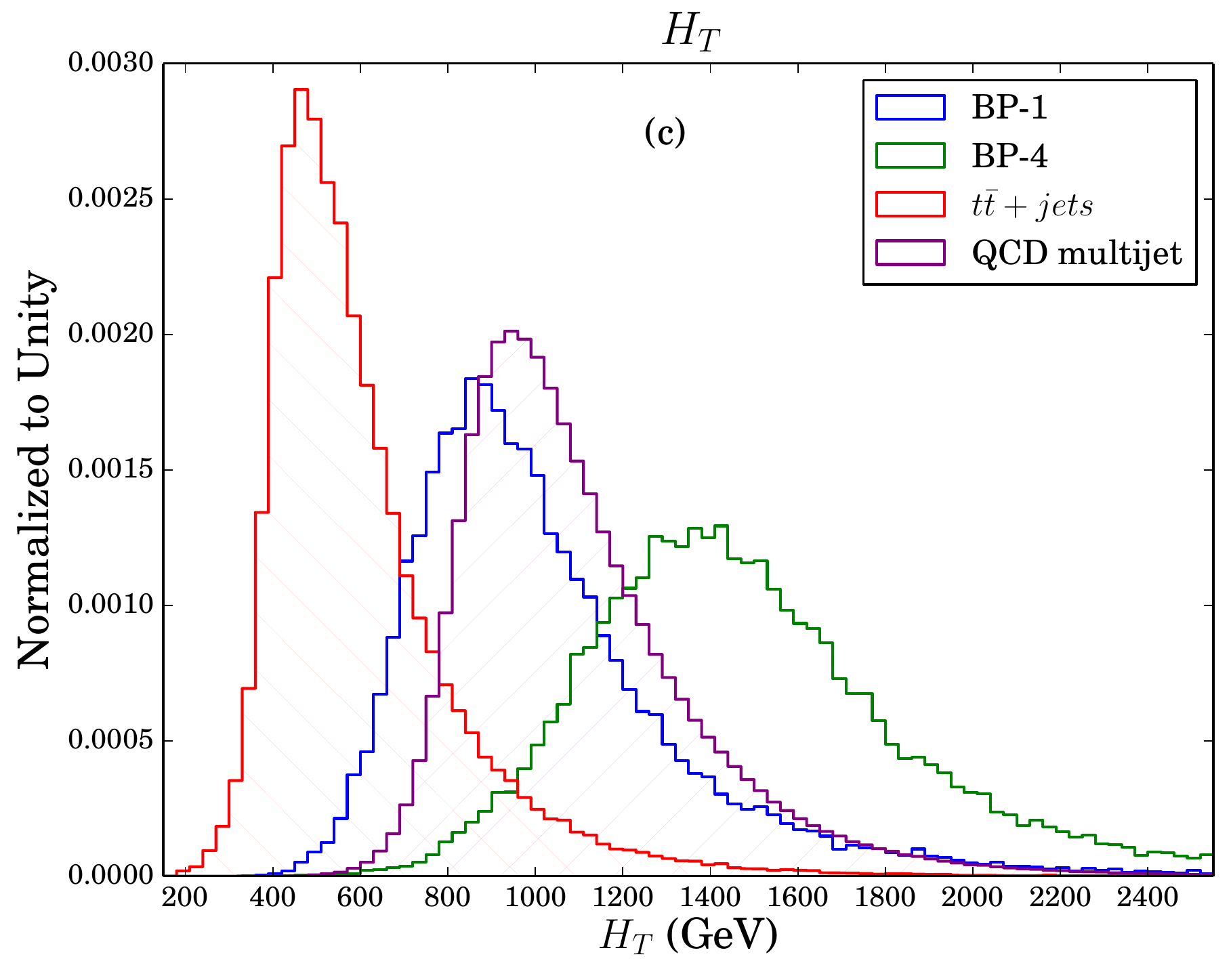}
\caption{{\small {\it The plots for a few of the variables used in the analysis. The left plot $(a)$ shows 
the $p_T$ of the hardest light jet (non top-tagged, non b-tagged jet), while the plot in the middle $(b)$ shows 
the $p_T$ of the hardest top-tagged jet. The rightmost figure $(c)$ is that of the $H_T$, which is 
the scalar sum of all the jets. In all of these, only two benchmark points (BP-1 and BP-4) have been 
shown and the histograms for the background processes are hatched.}}}
\label{fig:boost_vars}
\end{figure}

Not only are the light-jets in the signal sample often harder
  than those in the background, an analogous statement also holds for
  the respective top-jet constituent (especially for heavier
  sbottoms).  To utilize these characteristic differences, between
the signal and background events, we
consider the $p_T$ of the hardest top jet and the $p_T$ of the hardest
and second hardest light jets as BDT input variables. Like the
leptonic analysis, one of the most important variables is $H_T$ (see
eqn.\ref{HT_defn}), with the sum, obviously, running over all the
jets. Being closely associated with the
center-of-mass energy of the process, it too
is an important discriminator. It is important to remind our readers here that we use
only trimmed jets to construct the jet observables.  We use the number
of b-tagged jets as a discriminator by passing it as a variable for
MVA, as QCD decreases vastly if a b-tag is demanded. We could, instead, have 
put a cut on it
before the MVA---a pre-MVA cut---but as this would decrease the
background a lot, making the BDT analysis somewhat unreliable, 
we desist. In Fig.~\ref{fig:boost_vars}, we plot the distributions
in the $p_T$ of the hardest light jet, that of the hardest top tagged
jet and $H_T$. The QCD multijet sample was generated with an imposed
cut of 1~TeV on the $H_T$ and after demanding that the two hardest
jets in the sample be harder than 100~GeV. With the center of mass energy of the sbottom pair
production process being $\sim 1$ TeV, this ensures ample, yet relevant statistics for
the QCD multijet process. The variable $H_T$
turns out to be a good discriminator as the peak of higher mass
benchmarks lies to the right of the QCD peak, the tail of the
distribution only contributes to the signal peaks.

\begin{figure}[htb!]
\centering
\includegraphics[scale=0.4]{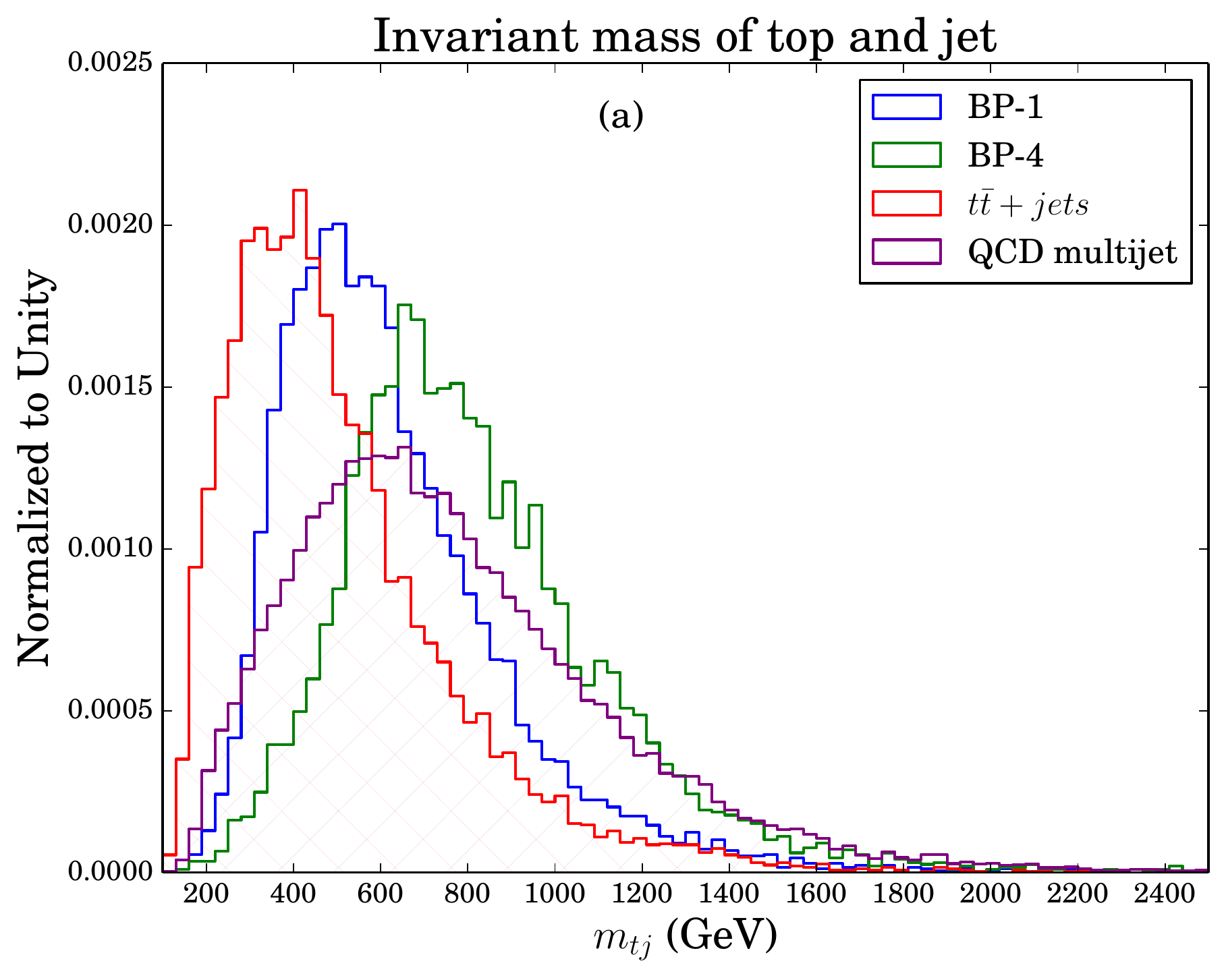}
\includegraphics[scale=0.4]{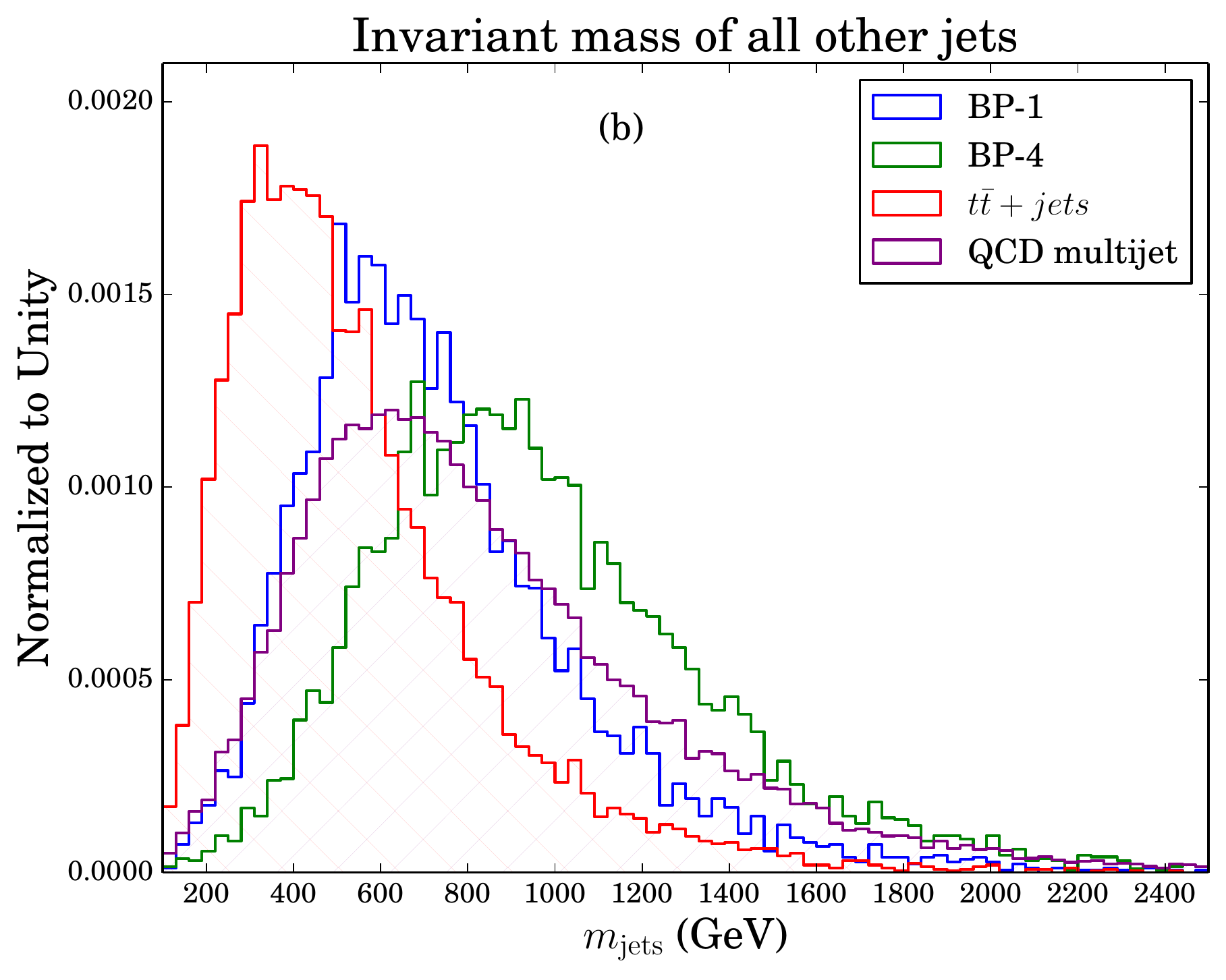}
\caption{{\small{\it The reconstructed masses that we used for the multivariate analysis. The plot on the left $(a)$ is 
the invariant mass of the top with one of the light jets (in short, `tj' set), while that 
on the right $(b)$ is the invariant mass of all the other jets in that event which do not correspond to 
the `tj' set. The invariant mass reconstruction technique has been discussed in the text in detail. 
Only two benchmarks (BP-1 and BP-4) are plotted and the background histograms are hatched.}}}
\label{fig:boost_recons_vars}
\end{figure}

\begin{figure}[htb!]
\centering
\includegraphics[scale=0.5]{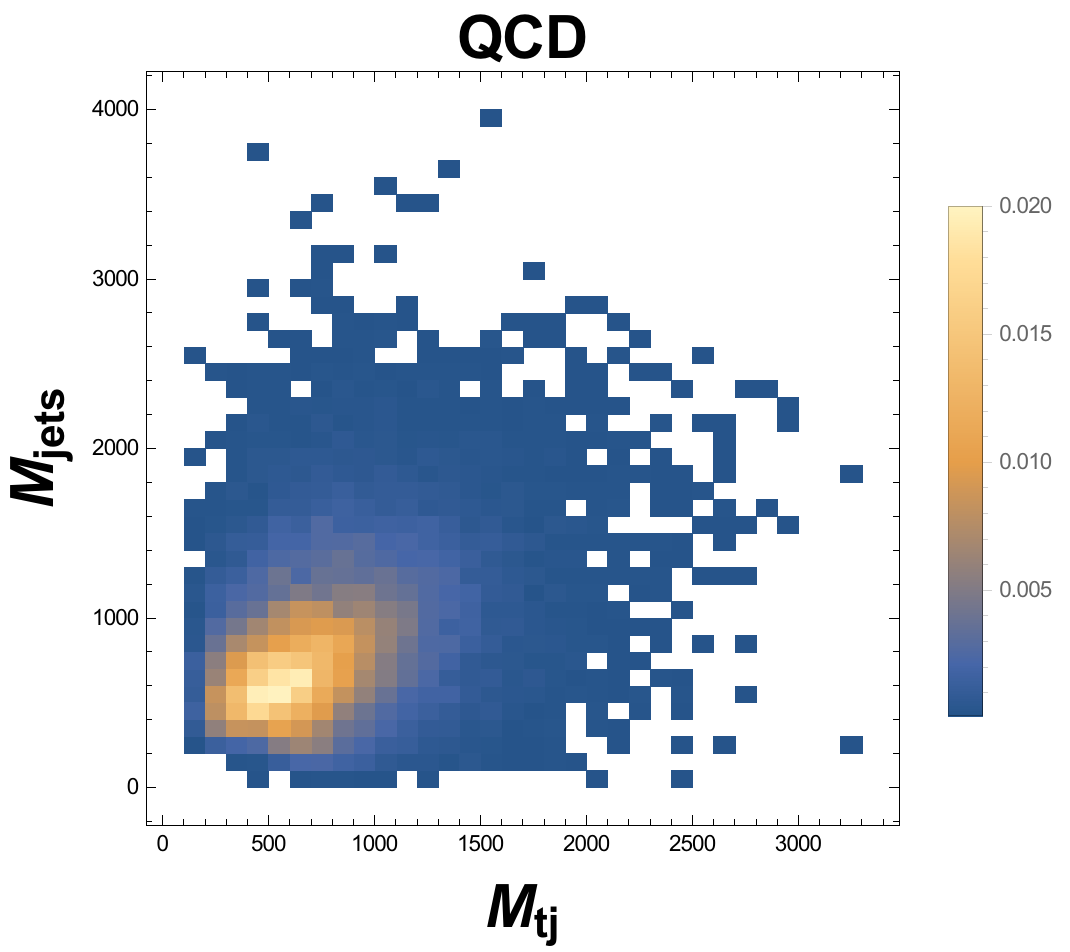}
\includegraphics[scale=0.5]{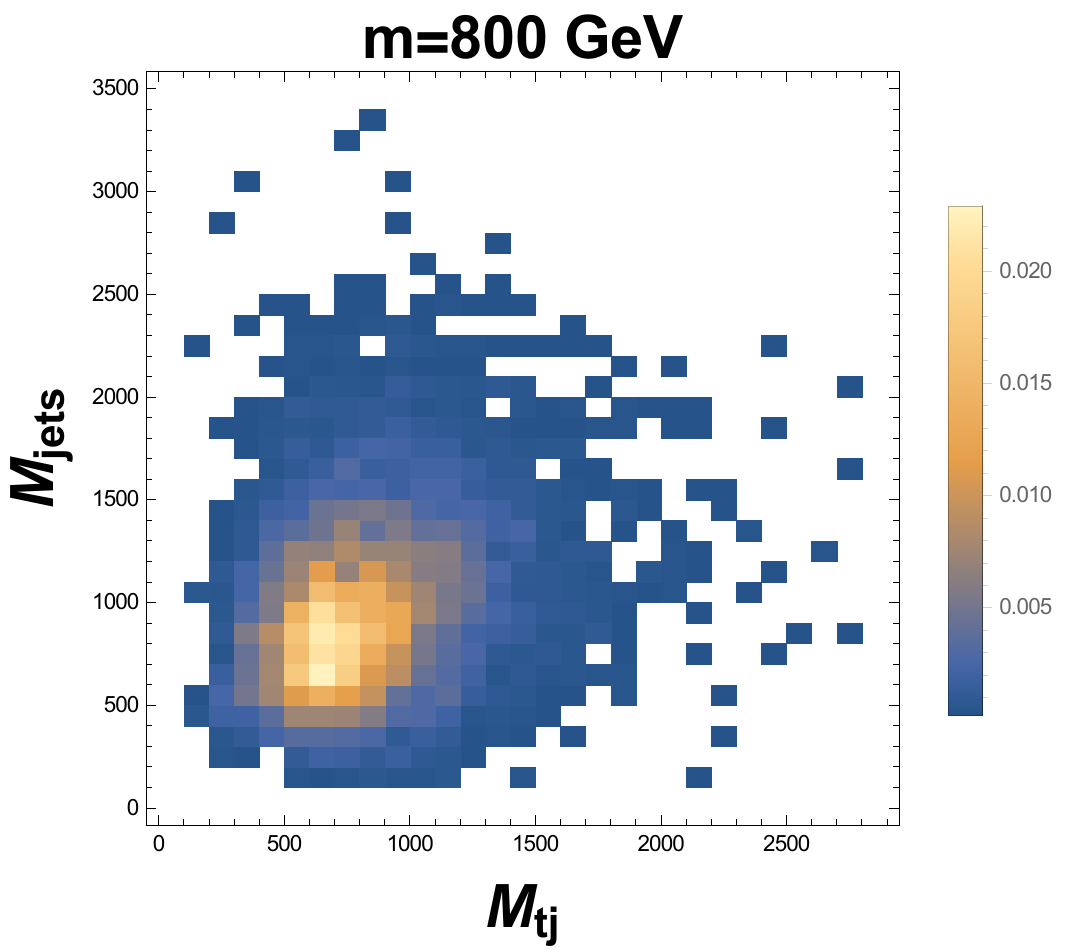}
\includegraphics[scale=0.5]{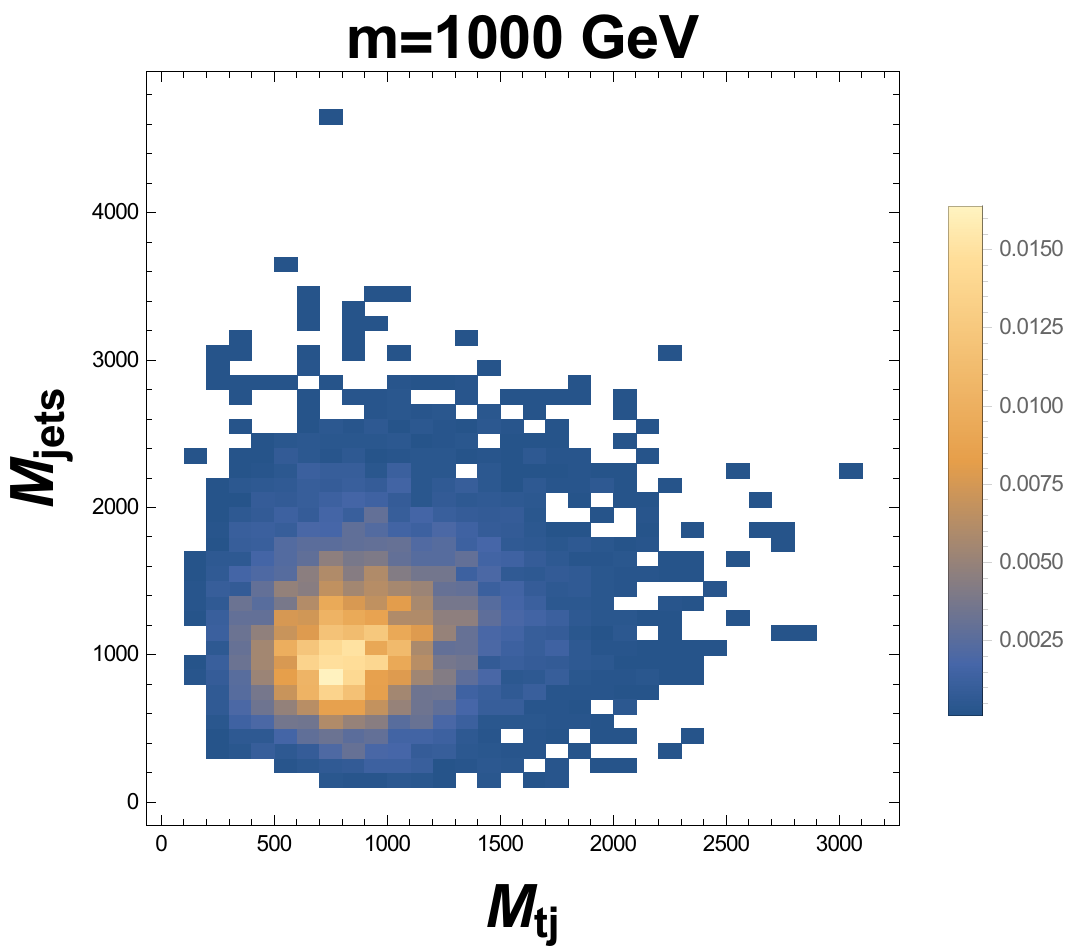}
\caption{{\small{\it Colour plots showing correlation between the $m_{tj}$ and $m_{\rm jets}$
variables in the case of QCD (left), BP-4 (middle) and BP-6 (right). Strong correlation can
clearly be seen around the (500, 500) point for QCD, (800, 800) for BP-4 and (1000, 1000) for 
BP-6. This can be exploited to probe heavier sbottoms as well.}}}
\label{fig:corr}
\end{figure}

Restricting ourselves to events with a tagged-top and at least four jets 
overall\footnote{Our primary event selection criteria includes at least 
one toptagged jet, however for the reconstruction of invariant 
masses we restrict ourselves to exactly one toptagged event. 
In principle, two or more toptagged samples would give better mass peaks 
with negligible QCD events, however we find a very few signal 
events surviving the two or more toptagged jet selection criteria.} 
two more useful observables are obtained by partitioning an event 
such that one subset contains the tagged top and a single non-b jet, 
while the other contains the rest of the jets. Denoting the invariant 
masses of the two sets by $m_{tj}$ and $m_{\rm jets}$, we retain these 
variables for the pairing that minimizes the difference
\begin{equation}
\Delta M \equiv | m(j_t,j_i) - m(j_k, j_l, ... ) |
\end{equation} 
Ideally, $\Delta M$ should vanish. However, owing to the 
vagaries of jet reconstruction algorithms as well as detector effects, 
this would rarely occur. Note that the requirement of the top's partner 
above being a non-b jet helps get rid of significant amount of the QCD 
background in the signal peak region. Note that, among the two invariant 
masses $m_{tj}$ and $m_{\rm jets}$ and the mass difference $\Delta M$, only 
two are independent parameters. In the MVA analysis, however, we use 
all the three parameters simultaneously as BDT inputs. In 
Fig.~\ref{fig:boost_recons_vars}, we plot the
two invariant masses we talked about earlier. These seem to have
moderate discriminatory powers.  Furthermore, In Fig.~\ref{fig:corr},
we also show the correlations in the $m_{tj} - m_{\rm jets}$ plane for
the QCD multijets (left plot), and for two benchmark points - BP-4
(middle) and BP-6 (right). For the QCD, the points are dense in the
region around the (500,500) point while for the signal it is dense
around (800,800) for BP-4 and around (1000,1000) for BP-6. It is
interesting to note that this feature, in principle, can be used for
probing heavier bottom squarks.

Nsubjettiness \cite{Thaler:2010tr} is an inclusive jet shape variable
which takes into account the energy distribution within a fat jet. It
is defined as 
\begin{equation}
\tau_N = \frac{\sum_k p_{T,k} {\rm min} \left(\Delta R_{1,k}, \Delta R_{2,k} ... \Delta R_{N,k}\right)}{\sum_k p_{T,k} R_0}
\end{equation}
where $\Delta R_{j,k}$ is the angular separation between the j$^{th}$
candidate jet and the k$^{th}$ constituent particle, $p_{T,k}$ is the
$p_T$ of the k$^{th}$ constituent and $R_0$ is the jet radius of the
fatjet under consideration.  Normalisation ensures
that $0 \leq \tau_N \leq 1$.  If $\tau_N \approx 0$, it indicates that
all the radiation in the jet is aligned with the subjet directions and
that there is a maximum of $N$ subjets in the considered jet. On the
other hand, if $\tau_N \gg 0$, it indicates the presence of more
subjets and that the radiation is distributed far from the candidate
subjets. It turns out that the ratio between two Nsubjettiness
variables might have higher discriminatory power than the variables
themselves. For events with at least one toptagged jet, we calculate
three such ratios, $\tau_{21}$, $\tau_{31}$ and $\tau_{32}$, where
$\tau_{ij} = \tau_i/\tau_j$, associated to the leading toptagged
jet. The ideal toptagged jet should have a three-prong structure, and
thus $\tau_{31}$ and $\tau_{32}$ are expected to be small, while this
would not be true for the QCD background. Thus, these can be used as good
discriminating variables. In Table~\ref{tab:boost_varlist}, we list
all the above-mentioned variables that are passed to the BDT for the
MVA.

Two further observables, namely $\rho$ and 
$\Phi(t,j)$, are used as BDT inputs, with the former being defined as  
\begin{eqnarray}
\rho &=& \frac{(p_T)_{j_t^{(1)}}}{(p_T)_{j_\ell^{(1)}}} 
\end{eqnarray}
where $(p_T)_{j_t^{(i)}}$ is the $p_T$ of the $i^{th}$ top jet, while
$(p_T)_{j_\ell^{(i)}}$ is the $p_T$ of the $i^{th}$ light jet (i.e. a
non top-tagged, non b-tagged jet). The quantity $\Phi(t,j)$ measures
the azimuthal angular separation between the top-tagged jet and the
leading light jet. 

\begin{savenotes}
\begin{table}
\centering
\begin{tabular}{|rl|l|}
\hline
 & {\bf Variable} & {\bf Definition} \\
\hline
1. & nlJet & The number of light jets in the event \\
2. & nbJet & The number of b-tagged jets in the event \\
3. & ntJet & The number of top-tagged jets in the event \\
4. & $(p_T)_{j_1}$ & $p_T$ of the hardest light jet \\
5. & $(p_T)_{j_2}$ & $p_T$ of the second hardest light jet \\
6. & $(p_T)_{j_t^{(1)}}$ & $p_T$ of the hardest top tag jet \\
7. & $H_T$ & scalar sum of the $p_T$ of all the jets \\
8. & $m_{tj}$ & the invariant mass of the top and jet system \\
9. & $m_{\rm jets}$ & the invariant mass of all the other jets \\
10. & $\Delta M = |m_{tj} - m_{\rm jets}|$ & the mass difference of the two reconstructed invariant masses \\
11. & $\tau_{21} = \tau_2/\tau_1$ & Ratio of the Nsubjettiness variables \\
12. & $\tau_{31} = \tau_3/\tau_1$ & Ratio of the Nsubjettiness variables \\
13. & $\tau_{32} = \tau_3/\tau_2$ & Ratio of the Nsubjettiness variables\\
14. & $\rho = \frac{(p_T)_{j_t^{(1)}}}{(p_T)_{j_\ell^{(1)}}}$ & Ratio of the hardest top-jet $p_T$ and light jet $p_T$ \\
15. & $\Phi(t,j)$ & Azimuthal angle separation between the toptagged jet and the leading light jet. \\
\hline
\end{tabular}
\caption{{\small {\it List of all the variables used in the multivariate analysis. Note that, the variable 
$\rho_2$ is calculated only only for events with two or more tagged tops.}}}
\label{tab:boost_varlist}
\end{table}
\end{savenotes}
\begin{table}
\centering
\begin{tabular}{|l|c|c|c|c|c|c|c|c|}
\hline
& { QCD} & {$ t \bar{t} +{\rm jets}$} & {BP-1} & {BP-2} & {BP-3} & {BP-4} & { BP-5} & { BP-6} \\
\hline
$\sigma_0$ (fb) & $1.9 \times 10^7$ & $8.3 \times 10^5$ & $5.2 \times 10^2$ & $1.8 \times 10^2$ & $6.7\times 10^1$ & $2.8 \times 10^1$ & $1.3 \times 10^1$ & $6.2$ \rule{0pt}{2.6ex}\\
$\sigma_{\rm toptag}$ (fb) & $2.6 \times 10^6$ & $6.7 \times 10^4$ & $1.4 \times 10^2$ & $5.0 \times 10^1$ & $2.0 \times 10^1$ & $8.6$ & $4.0$ & $2.0$ \\
\hline
\end{tabular}
\caption{{\small{\it Showing the initial cross-section ($\sigma_0$) and 
surviving cross-section after at least one top is tagged 
($\sigma_{\rm toptag}$) for the background and all the signal benchmarks. The 
QCD multijet sample is generated after a cut on the $H_T$ variable of 800 GeV 
and cut of 100 GeV on the $p_T$ of the two hardest jets.}}}
\label{tab:had_xsec}
\end{table}

\subsection{Results}

\begin{figure}[htb!]
\centering
\includegraphics[scale=0.5]{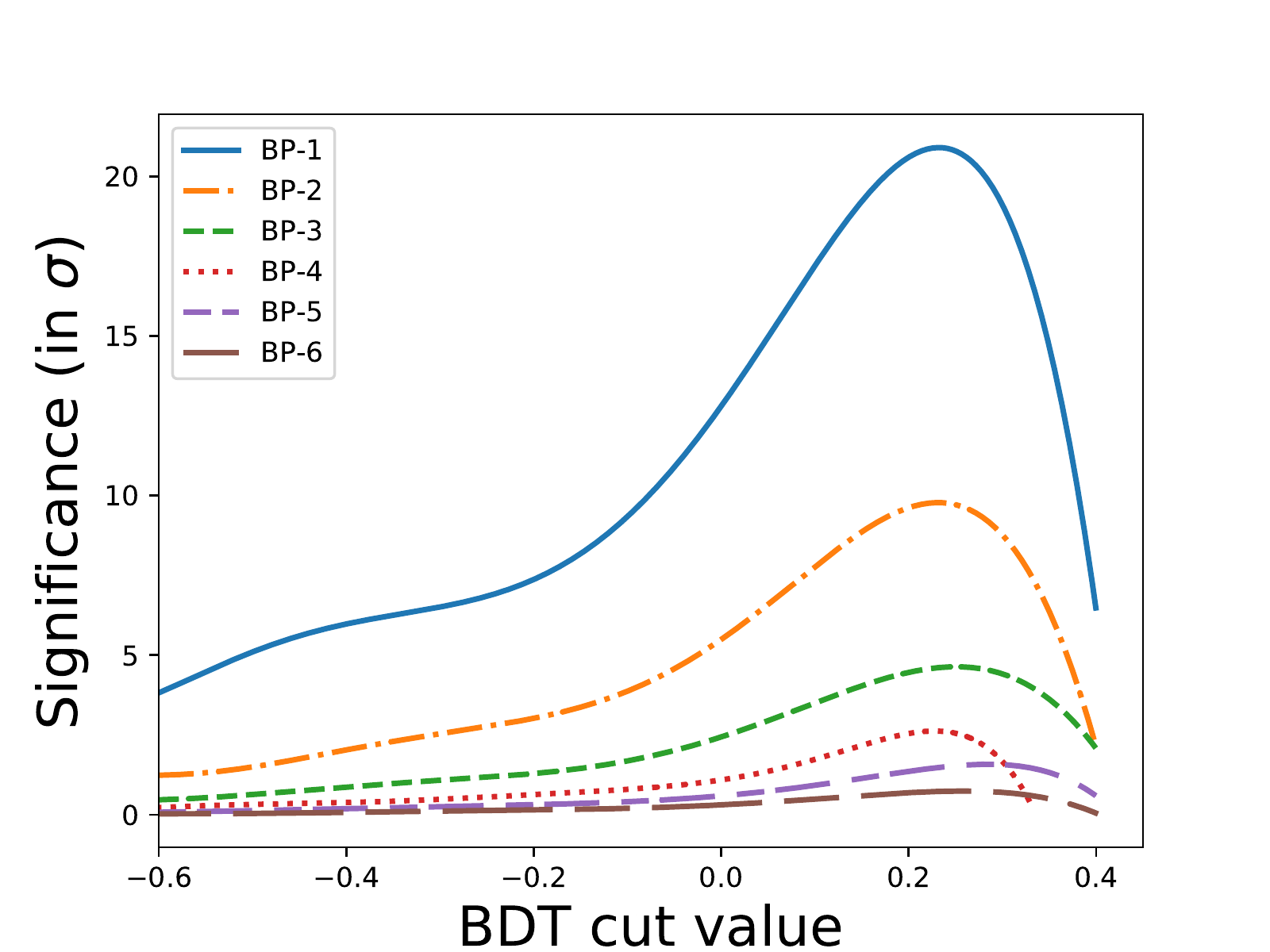}
\caption{{\small{\it Plot showing the significance of the different benchmark points with variation
of the BDT cut value. All figures for for 300 ${\rm fb}^{-1}$}}}
\label{fig:sig_had}
\end{figure}

We now proceed to discuss the details of the multivariate analysis
using the BDT method implemented in the TMVA ROOT framework. The
fifteen variables discussed earlier and listed in
Table~\ref{tab:boost_varlist}), each of which we expect to have some
discrimination power, are used as the 
BDT
inputs. The BDT parameters are same as in the leptonic case,
viz. {\ NTrees =} 400, {\tt MaxDepth =} 5 and {\tt MinNodeSize =}
2.5\% with {\tt AdaBoostBeta =}~0.5. In Table~\ref{tab:had_xsec}, we
show the initial cross-sections $(\sigma_0)$ and the cross-section
after at least one top is tagged ($\sigma_{\rm toptag}$). The top
tagged events in the QCD samples are due to misidentification of fat
jets as top jets; the 
corresponding `fake rate' is about
10\% for the QCD sample. The advantage of using the multivariate
  analysis is that we can translate a complicated multi-dimensional
  optimisation problem over all input variables into that involving a
  one parameter function which is much easier to handle. We can now
  choose the BDT cut value such that it maximizes the signal
  significances. The results of the multivariate analysis are
  presented in Table \ref{tab:boostbdtres} and
  Fig.~\ref{fig:sig_had}.

\begin{table}[htb!]
\centering
\begin{tabular}{|c|cccccc|}
\hline
 & BP-1 & BP-2 & BP-3 & BP-4 & BP-5 & BP-6\\
\hline
\hline
$m_{\tilde{b}_1}$(GeV) & 500 & 600 & 700 & 800 & 900 & 1000 \\
\hline
BDT cut & 0.186  & 0.167 & 0.245 & 0.238 & 0.266 & 0.280 \rule{0pt}{2.4ex} \\
$\mathcal{S} = \frac{S}{\sqrt{S+B}}$ & 3.50 & 1.21 & 0.57 & 0.32 & 0.20 & 0.16  \\
\hline
\end{tabular}
\caption{{\small{\it Signal significances for 
the benchmark points with the choice of BDT cuts with 
$\mathcal L = 300~{\rm fb^{-1}}$ of integrated luminosity.}}}
\label{tab:boostbdtres}
\end{table}

From Table~\ref{tab:boostbdtres}, it is evident that the signal
significances for the benchmark points diminish rapidly as we proceed
from BP-1 to BP-6. The primary reason is the rapid decrease of the
sbottom pair production cross-section with the increase of sbottom
mass. Even though we expect to tag the top quarks originating from the
heavier sbottoms more efficiently, the impact is negligible compared
to the drastic fall in the production cross-section.  Improved
toptagging with smaller fake rate for the QCD multijet events is
essential for better signal-background
discrimination. State-of-the-art jet grooming techniques, namely
Pruning \cite{Ellis:2009su} and SoftDrop \cite{Larkoski:2014wba} may
help in reducing QCD multijet events and, thus, 
enhancing the signal significance.

It is thus clear that the hadronic channel is not the most favourable
one for the discovery of the sbottom. The
exclusion limit is barely reached, for 300 fb$^{-1}$
integrated luminosity, for the first benchmark point (500
GeV). A higher integrated luminosity of 3000 fb$^{-1}$ from a futuristic collider
like the High Luminosity LHC will be able to push the exclusion limit
to about 650 GeV. However, here we would like to suggest an
interesting extension of this analysis which combines our leptonic and
hadronic analyses, a successful marriage of the boosted and
non-boosted analyses with semi-leptonic final states. We will leave
this very interesting avenue for our future work.

\section{Summary and Outlook}
In this paper, we analyze the discovery potential of the LHC for a
bottom-squark (the LSP) which decays, with a $100\%$ branching ratio,
to the top
and a light quark via $R-$parity violating $UDD$ couplings. While 
relatively heavy squarks allow for large couplings, thereby 
opening up the possibility of significant resonance production (such as 
$d + b \to \tilde t^*$), we eschew this possibility altogether, 
assuming the couplings are small enough for them to be unimportant 
in production processes (whether resonance or pair), yet large enough to 
preclude recognizably displaced vertices. 

Based on
the final state, we devise two strategies, one for a final state which
has at least one isolated lepton (electron or muon) and the other for 
a fully hadronic final
state. For the leptonic state, two independent investigations have been
performed: first, using the traditional
cut-based analysis and then using a multivariate analysis (MVA). 
The backgrounds
considered for the leptonic analysis are $t\bar {\rm + jets}$, $t \bar{t} b
\bar{b}$, $t \bar{t} W$, $t \bar{t} H$ and $t \bar{t} Z$. After
demanding an isolated lepton tag, we consider cuts on various observables, like
$H_T$ and $M_{T2}$ among others, in order to separate signal from
background. We also reconstruct the sbottom mass with exactly one 
isolated lepton with two or more b-tagged jets and four hardest light 
({\em i.e.}, non-b-tagged) jets in the event. We use these reconstructed sbottom masses, namely 
$m^{h}_{\tilde{b}_1}$ and $m^{\ell}_{\tilde{b}_1}$ representing the reconstructed masses 
using the hadronically and leptonically decaying top quarks respectively, as BDT inputs.
While the cut-based analysis reveals an exclusion of $\sim 750$~GeV of the sbottom mass, 
the MVA extends that range to $\sim 850$~GeV with 300 ${\rm fb}^{-1}$ of data.

For the fully hadronic final state, we perform the MVA directly as we
find through our leptonic analysis that it helps to improve the reach
for heavy resonance. The dominant QCD multijet and the ($t \bar t +$
jets) backgrounds drown out all other sources of SM backgrounds. We
consider events in which we can tag at least one top jet using the
{\tt HEPTopTagger} framework. Furthermore, in order to reduce the
effect of underlying events and soft radiation, we groom the large R
anti-kT jets using the ``Trimming" technique. Several observables are
then constructed using these trimmed jets and then passed to the
MVA. The results, unfortunately, are not as good as in the leptonic
channel, with the exclusion limit barely crossing $500$~GeV.
 
The sensitivity that our analyses project can be further improved upon
the inclusion of other aspects. We list a few here:
\begin{itemize}
\item Incorporating tracker information such as number of soft-tracks \cite{Chakraborty:2016qim},
  not associated with the reconstructed objects, is likely to help in
  improving the sensitivity, especially for the most challenging case,
  viz.  the fully hadronic final state.  

\item As we have already mentioned, the very couplings that we have 
  investigated here also lead to the stop decaying to a bottom 
  and a light quark. Although the background processes to the corresponding
  final state (two b + d/s pairs) have larger production rates than the 
  one here, the
  simpler nature of the final state, especially the ability to 
  reconstruct the masses~\cite{Choudhury:2005dg,Choudhury:2011ve} 
  allows for a higher experimental sensitivity~\cite{ATLAS:2016yhq}. 
  In this work, we have deliberately avoided this channel, assuming the 
  stop to be much heavier. If it is not so, but is comparable to the 
  sbottom in mass, the sensitivities need to be compounded.

\item The very same coupling will also lead to decays like $\tilde d
  \, (\tilde s) \to \bar t + \bar b$ (depending on the identity of the
  coupling). Once again, we have not included this assuming that the
  $\tilde d \, (\tilde s)$ is much heavier. This assumption was partly
  motivated by the need to keep large FCNCs at bay.  However, a second
  solution exists if the squark masses are relatively
  degenerate~\cite{Choudhury:1994pn,Misiak:1997ei,Brignole:1997dp}. This
  can be motivated if the soft-supersymmetry breaking masses for the
  right-handed squarks are similar, and so are the small trilinear
  terms $A_d, A_s, A_b$. As can be readily appreciated, this solution
  is more natural than the one we have considered here.

  Direct two-body decays of $\tilde d \, (\tilde s)$ that are nearly
  degenerate with the $\tilde b$ would lead to configurations very
  similar to that we have considered here, with the added advantage
  that the non-top jets here originate from $b$-quarks and, thus, can
  be tagged.  This would severely curtail the SM backgrounds (with the
  biggest effect being seen in the fully hadronic state), resulting in
  much improved sensitivity.

\item Indeed, even if the $\tilde d \, (\tilde s)$ are sufficiently heavier
 than the $\tilde b$ (on account of a possibly large $A_b$, the
 effects due to $A_{d,s}$ of similar magnitudes being smaller) so as
 to open up their R-conserving three-body decays (into $\tilde b$
 accompanied by a pair of quarks), the associated quarks would lead to
 only soft jets. Thus, for such a cascade, one essentially comes back
 to the configuration that we have analysed here.
    
\item Finally, both stops and sbottoms (and, similarly, the
  other squarks) can originate from gluinos. If the gluino is
    not much heavier than the quark, its production cross section is
    much larger. Such a gluino would decay into the squark-quark
    pair. The latter would lead to a soft jet, with the first
    suffering a R-violating decay leading to a configuration very
    similar to the one under consideration.  This is quite analogous
    to the ATLAS study~\cite{ATLAS:2016nij} (that set a limit of
    $m_{\tilde g} > 1.08$ TeV), except for the fact that, in the
    present context, some of the jets would be rather soft.

  On the other hand, if the gluino is very heavy,
  the produced squarks will be highly boosted,
  providing highly boosted tops in turn. This suits top tagger
  algorithms favorably and has been analyzed in
  \cite{Bhattacherjee:2013tha}. 

\end{itemize}
In view of these obvious improvements to the sensitivity that 
can be effected, it is quite apparent that the conclusions reached 
by us are only conservative. 
\vskip 1.5cm

\small
\baselineskip 16pt
\noindent
{\large \bf {Acknowledgments:}}\\
AC and DB wish to thank Gouranga Kole, 
Soureek Mitra, Shankha Banerjee and Seema Sharma for 
helping us resolve various ROOT and TMVA-related issues. 
AC and DB would also like to thank Sabyasachi Chakraborty for
many fruitful discussions. DC acknowledges partial support from the European Union's Horizon 2020
research and innovation program under Marie Sklodowska-Curie grant
No 674896. The computations reported here were 
performed on the computational resources of the Department of 
Theoretical Physics, TIFR. DB thanks Department of Theoretical Physics, IACS for
hospitality during the completion of this work.

\end{document}